\documentclass[%
reprint,
superscriptaddress,
amsmath,amssymb,amsthm,mathrsfs,
aps,
prl,
]{revtex4-1}

\usepackage{graphicx}
\usepackage{dcolumn}
\usepackage{bm}
\usepackage[usenames]{color}
\usepackage[ 
margin=0.75in,
]{geometry}
\usepackage[normalem]{ulem}

\graphicspath{{./figs/}}

\def\Tmit{$T_{\mathrm{MIT}}$}

\def\Rw{$R_w$}

\def\vo{V$_2$O$_3$}
\def\dof{d.o.f.}
\def\qmax{$Q_{\mathrm{max}}$}

\begin{document}
	
	\title{
		Uniform structural phase transition in \vo\ without short-range distortions of the local structure
	}
	
	\author{Ethan R. A. Fletcher}
	\affiliation{ %
		Department of Physics and Astronomy, Brigham Young University, Provo, Utah 84602, USA.
	} %

	\author{Kentaro Higashi}
	\affiliation{ %
		Department of Energy and Hydrocarbon Chemistry, Graduate School of Engineering, Kyoto University, Nishikyo, Kyoto 615-8510, Japan
	} %
	
	\author{Yoav Kalcheim}
	\affiliation{ %
		Department of Materials Science and Engineering, Technion - Israel Institute of Technology, Haifa 32000, Israel
	} %
	
	\author{Hiroshi Kageyama}
	\affiliation{ %
		Department of Energy and Hydrocarbon Chemistry, Graduate School of Engineering, Kyoto University, Nishikyo, Kyoto 615-8510, Japan
	} %
	
	\author{Benjamin A. Frandsen}
	\affiliation{ %
		Department of Physics and Astronomy, Brigham Young University, Provo, Utah 84602, USA.
	} %

	\begin{abstract}
		The local structure of V$_{2}$O$_{3}$, an archetypal strongly correlated electron system that displays a metal-insulator transition around 160~K, has been investigated via pair distribution function (PDF) analysis of neutron and x-ray total scattering data. The rhombohedral-to-monoclinic structural phase transition manifests as an abrupt change on all length scales in the observed PDF. No monoclinic distortions of the local structure are found above the transition, although coexisting regions of phase-separated rhombohedral and monoclinic symmetry are observed between 150~K and 160~K. This lack of structural fluctuations above the transition contrasts with the known presence of magnetic fluctuations in the high-temperature state, suggesting that the lattice degree of freedom plays a secondary role behind the spin degree of freedom in the transition mechanism.
	\end{abstract}
	
	\maketitle
	
	\section{Introduction}	
	The interplay among spin, orbital, and lattice degrees of freedom (\dof) in solids has long commanded a large share of research efforts in condensed matter physics. Rich and diverse behaviors are observed in systems where these \dof\ are simultaneously active, including metal-insulator transitions (MITs), high-transition-temperature superconductivity, colossal magnetoresistance, unconventional magnetism, and novel forms of symmetry breaking~\cite{imada;rmp98,keime;np17}. Transition metal oxides in particular have been the subject of intensive investigations in this direction, thanks to the strong electronic correlations and the unfilled $d$ shell that naturally leads to interactions between the spin, orbital, and lattice \dof~\cite{tokur;s00}
	
\begin{figure}
		\includegraphics[width=65mm]{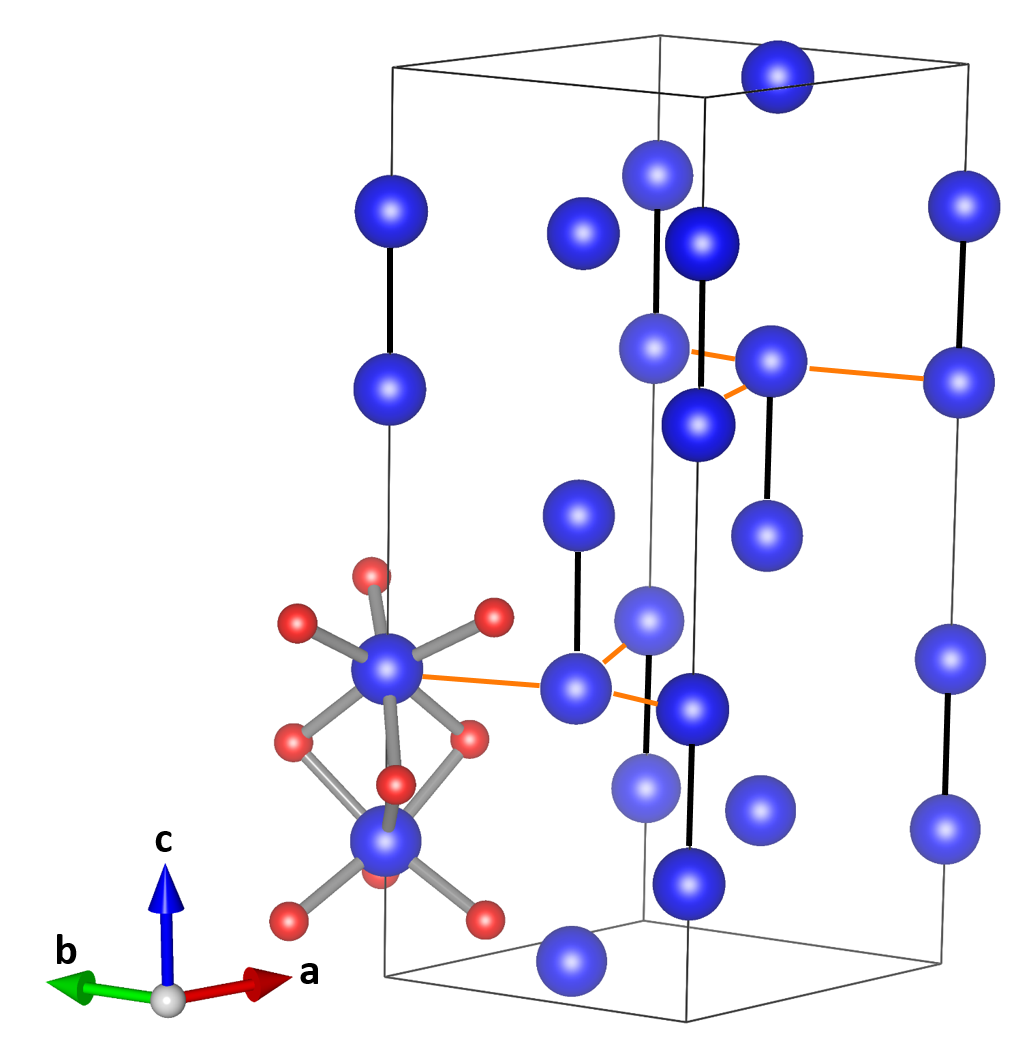}
		\caption{\label{fig:structure} Rhombohedral crystal structure of \vo. Blue and red spheres represent vanadium and oxygen atoms, respectively. Octahedrally coordinated oxygen atoms surround each vanadium atom, but for simplicity, the oxygen octahedra are shown only for two vanadium atoms. The thick black (thin orange) lines connect nearest-neighbor (next-nearest-neighbor) vanadium atoms.}
	\end{figure}

	Among transition metal oxides, \vo\ has occupied a prominent position in the study of MITs in Mott insulators and other strongly correlated electron systems~\cite{mcwha;prl69,mcwha;prl71,mcwha;prb73}. At \Tmit\ $\sim$155~K on cooling and 170~K on warming, \vo\ transitions between a paramagnetic metal with a rhombohedral crystal structure above \Tmit\ and an antiferromagnetic insulator with a monoclinic structure below \Tmit~\cite{derni;jpcs70,moon;prl70,uemur;hfi84,denis;jap85,bao;prl93,frand;nc16}. The high-temperature rhombohedral crystal structure is illustrated in Fig.~\ref{fig:structure}. The monoclinic distortion is characterized by a tilting of the nearest-neighbor vanadium pairs (connected by thick black lines in Fig.~\ref{fig:structure}) off the $c$ axis and a rotation of the associated oxygen octahedra. As a result, the three equivalent next-nearest-neighbor vanadium-vanadium bonds in the rhombohedral structure (orange lines in Fig.~\ref{fig:structure}) break into three distinct bonds with slightly different bond lengths. Simultaneously, the magnetic moments order ferromagnetically within (110) planes in the hexagonal setting, transforming to (010) planes in the monoclinic setting, with alternating magnetization between layers.
	
	The concomitant changes of the electronic, magnetic, and structural properties across the MIT underscore the complex nature of the transition and the nontrivial interdependencies among these various \dof\ In bulk \vo, the transition can be tuned by substituting small amounts of Ti or Cr for V, introducing V vacancies, or applying hydrostatic pressure, which results in a first-order quantum phase transition to a paramagnetic metallic state around 2~GPa~\cite{gossa;prb70,mcwha;prb73,ueda;jpcs78,frand;nc16}. Further possibilities for engineering the MIT can be achieved by fabricating \vo\ in thin-film form~\cite{grygi;apl07,ji;apl12,dille;apl14,kalch;afm20,lee;acsami21}. Although significant progress has been made toward a deeper understanding of this remarkable transition in \vo\cite{mcwha;prl69,caste;prb78a,caste;prb78b,caste;prb78c,paola;prl99,park;prb00,rozen;prl95,held;prl01,mo;prl03,kelle;prb04,kotli;pht04,hansm;pss13}, a complete theory that successfully describes all the observed behavior remains elusive.
	
	In recent years, elucidating the interrelationships between the electronic, magnetic, and structural phase transitions in \vo\ has been a major research objective~\cite{lupi;nc10,mcleo;np16,kalch;prl19,frand;prb19b,kundu;prl20,chen;aplm20,hu;prb21}. Particularly important is determining which \dof---i.e. spin, lattice, or orbital---is the primary driver of the transition. Experimental efforts in this direction have focused on determining if the electronic, magnetic, and structural transitions can be decoupled from each other and/or if fluctuations of any \dof\ are present in the high-temperature phase, which would provide insights into the hierarchy of interactions underlying the threefold transition in \vo. Conflicting viewpoints have emerged, leaving the question open. For example, a nanoscale-resolved infrared spectroscopy study suggested a decoupling of the structural transition from the metal-insulator transition~\cite{mcleo;np16}, while x-ray diffraction~\cite{kalch;prl19} and muon spin relaxation~\cite{frand;prb19b} studies with resistance-calibrated thermometry found that all three transitions remained tightly coupled. Antiferromagnetic fluctuations have been observed in the high-temperature state~\cite{bao;prl97,bao;prb98,frand;prb19b,trast;prb20}, suggesting the importance of the spin \dof\ On the other hand, extended x-ray absorption fine structure (EXAFS) studies of the local atomic structure have reported the presence of short-range monoclinic distortions in the nominally rhombohedral phase acting as a structural precursor to the MIT~\cite{frenk;ssc97,pfalz;prb06}, while sound velocity measurements have suggested a softening of certain elastic moduli on cooling toward the MIT~\cite{mulle;jap05,mulle;jap08,kunde;apl13,seikh;ssc06}. Reminiscent of numerous examples of local symmetry breaking and structural inhomogeneity preceding long-range structural transitions in strongly correlated materials~\cite{billi;prl96,qiu;prl05,dagot;s05,bozin;sr14,bozin;nc19,perve;nc19}, such a scenario would underscore the importance of the lattice \dof\ Additional experimental work is therefore required to clarify these important issues in \vo.
	
	In this work, we use x-ray and neutron pair distribution function (PDF) analysis to perform a detailed investigation of the local atomic structure of pure \vo\ at ambient pressure. The PDF method provides direct access to the local pairwise atomic correlations in real space via Fourier transformation of the scattering data, revealing details of the local structure that may otherwise be invisible to conventional diffraction techniques~\cite{egami;b;utbp12}. We find that the local atomic structure transitions abruptly and uniformly between the monoclinic and rhombohedral symmetries at \Tmit\ on all length scales in the PDF data, with no evidence for short-range monoclinic distortions surviving in the high-temperature phase. The lack of any structural fluctuations above the transition suggests that the lattice \dof\ is inactive in the high-temperature state, in contrast to the known presence of antiferromagnetic fluctuations at high temperature. These results strongly suggest that the lattice plays a secondary role in the MIT behind the spin degree of freedom.
	
	\section{Experimental Methods}
	A powder sample of V$_{2}$O$_{3}$ was synthesized via reduction of high-purity NH$_4$VO$_3$ in 5\% H$_2$/Ar gas at 900$^{\circ}$C with a flow rate of 3.3~ml/s for two days, followed by cooling at 100$^{\circ}$C/h. The magnetic susceptibility was measured using a SQUID magnetometer (Quantum Design, Magnetic Property Measurement System), revealing a sharp transition at 170~K on warming [see Fig.~\ref{fig:fits}(d)]. This confirms the good quality of the sample with minimal vanadium vacancies.
	
	The neutron PDF experiment was conducted at the Spallation Neutron Source (SNS) at Oak Ridge National Laboratory on the NOMAD beamline~\cite{neuef;nimb12}. Powder diffraction patterns were collected in a warming sequence between 100 and 400~K from 300~mg of powder in a thin quartz capillary using a nitrogen gas blower for temperature control. At each temperature, scattering data were collected for a total accelerator proton charge of 4~C. The data reduction and Fourier transform were performed using the data processing scripts on the beamline computers. The maximum momentum transfer \qmax\ included in the Fourier transform was 35~\AA$^{-1}$. The x-ray PDF experiment was performed on beamline 28-ID-1 of the National Synchotron Light Source II (NSLS-II) at Brookhaven National Laboratory using an x-ray wavelength of 0.1671~\AA\ on a 10~mg sample in a thin kapton capillary sealed with clay. The sample was placed in a liquid helium cryostat capable of a temperature range of 5 K to 500 K. The diffraction patterns were recorded on a large area detector made of amorphous silicon and azimuthally integrated using Fit2D~\cite{hamme;esrf04} to obtain one dimensional diffraction patterns. These diffraction patterns were normalized and Fourier transformed with $Q_{\mathrm{max}}=24$~\AA$^{-1}$ to produce the PDF using the xPDFsuite software available at the beamline. Diffraction images were collected in a warming sequence from 5~K to 400~K in steps of approximately 5~K, with two minutes of data collection per temperature point.
	
	Atomic PDF analysis was conducted using PDFgui~\cite{farro;jpcm07} and DiffPy~\cite{juhas;aca15}. Magnetic PDF~\cite{frand;aca14,frand;aca15,frand;prl16} analysis of the magnetic correlations in \vo\ was performed using the diffpy.mpdf package.
	
	\section{Results}
	
	We first present the atomic PDF analysis of the neutron total scattering data, followed by the analysis of the x-ray data. Neutrons scatter strongly from oxygen but only very weakly from vanadium, whereas x-rays scatter much more strongly from vanadium than from oxygen. Hence, these two data sets offer complementary sensitivity to oxygen (neutrons) and vanadium (x-rays). We then return to the neutron data set for the magnetic PDF analysis of the antiferromagnetic correlations in \vo.
	
	\subsection{Neutron PDF analysis}
\begin{figure}
		\includegraphics[width=75mm]{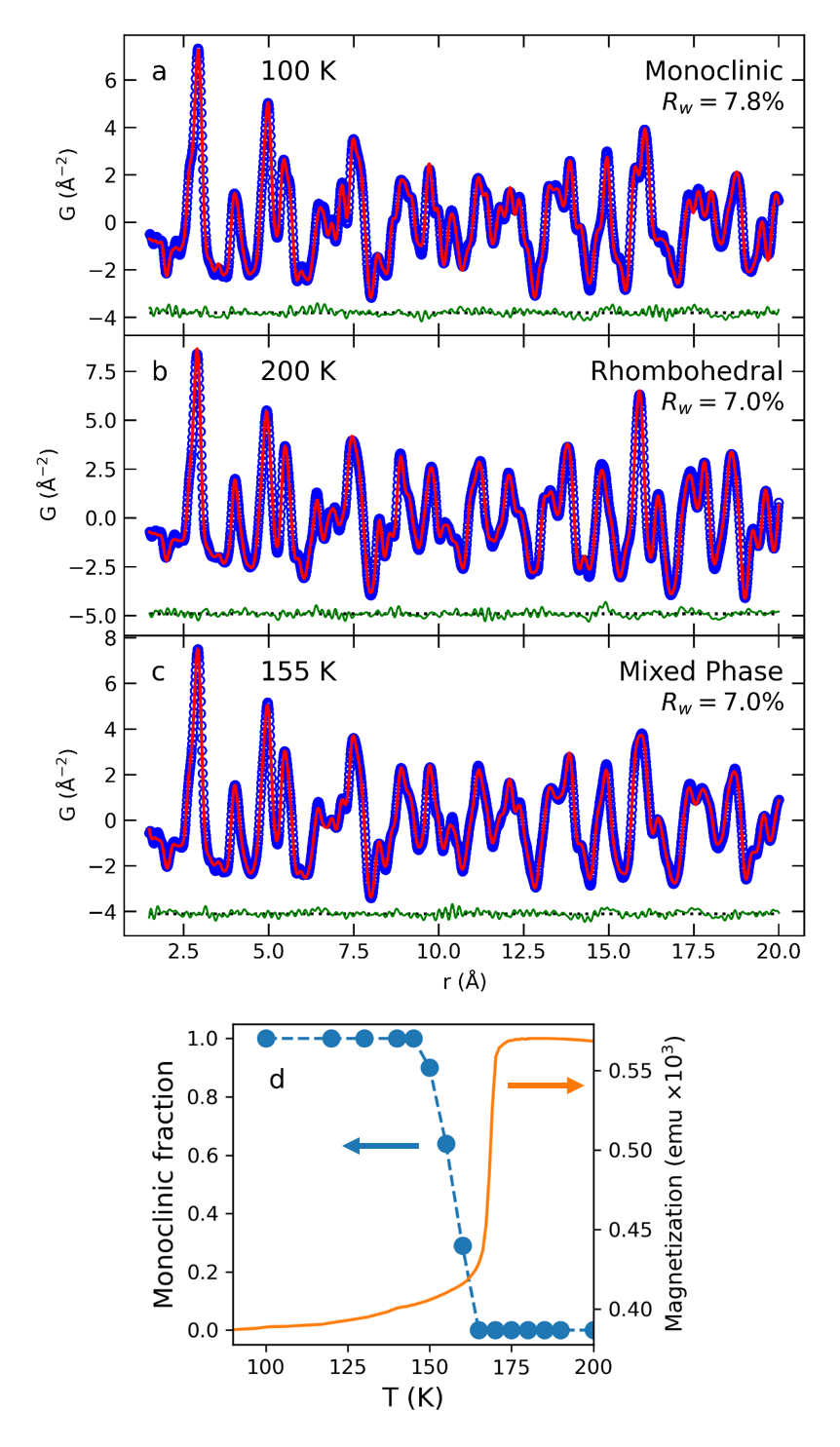}
		\caption{\label{fig:fits} Fits to the neutron PDF data collected at 100~K using the monoclinic structural model (a), 200~K using the rhombohedral model (b), and 155~K using a mixed phase model with spatially segregated regions with the rhombohedral and monoclinic structures (c). (d) Monoclinic phase fraction extracted from the PDF fits (blue circles, left axis) and magnetic susceptibility obtained from SQUID magnetometry (orange curve, right axis). Both measurements were done in a warming sequence.}
		
	\end{figure}

	In Fig.~\ref{fig:fits}, we display atomic PDF fits to the neutron PDF data collected at 200~K (a) and 100~K (b) using the published rhombohedral and monoclinic structural models, respectively.
	The fit quality is good, as evidenced by the low values of the goodness-of-fit metric $R_w = \sqrt{\frac{\sum{\left(G_{\mathrm{obs}}-G_{\mathrm{calc}}\right)^2}}{\sum G_{\mathrm{obs}}^2}}$, where $G_{\mathrm{obs}}$ and $G_{\mathrm{calc}}$ are the observed and calculated PDF patterns, respectively. This confirms that the sample is phase pure and undergoes a structural transition from rhombohedral to monoclinic symmetry, as expected. Additionally, we display a fit at 155~K conducted with a model containing both the rhombohedral and monoclinic phases, corresponding to phase separation of the two structures at this temperature. Fits using either only the rhombohedral model or the monoclinic model were unsatisfactory ($R_w > 12\%$, a much larger value of \Rw\ compared to the two-phase model and well beyond the conservative estimate of 0.8\% as the threshold for a meaningful difference between models~\cite{bird;jac21}). Smaller improvements to \Rw\ when including two phases are also seen for 150~K and 160~K, whereas single-phase fits are satisfactory outside this temperature range. Thus, the structural transition is in process from slightly below 150~K to slightly above 160~K. This type of structural phase separation has been observed previously in \vo~\cite{kalch;prl19,frand;prb20}, along with phase separation between paramagnetic and antiferromagnetic regions~\cite{frand;nc16,frand;prb20}. The monoclinic phase fraction extracted from the fits is shown as a function of temperature in Fig.~\ref{fig:fits}(d), together with the magnetization curve obtained from SQUID magnetometry. The results are consistent between the two probes, although the transition temperature observed in the PDF data appears to be $\sim$5~K lower than the transition temperature observed in the magnetometry data. This is likely due to imperfect temperature calibrations. 
	
	\begin{figure*}
		\includegraphics[width=150mm]{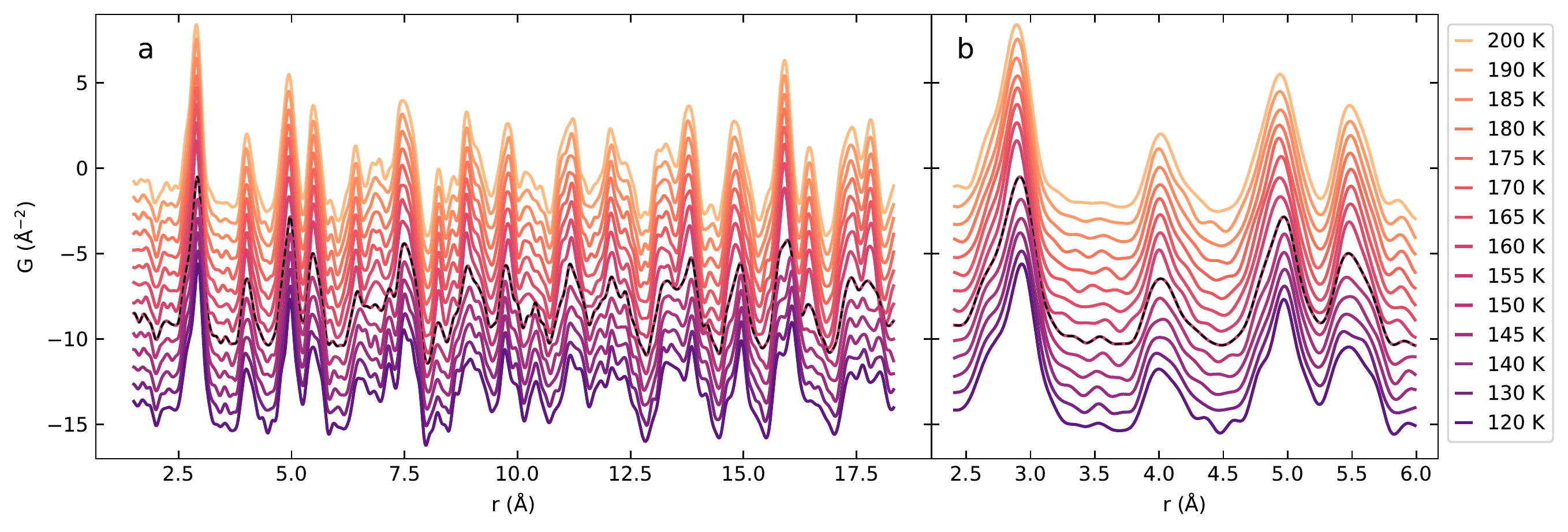}
		\caption{\label{fig:neutronwaterfall} Neutron PDF patterns for \vo\ at various temperatures across the transition shown over a wide viewing range (a) and a narrower range focusing on the local atomic correlations (b), showing an abrupt change in structure across the transition. The patterns have been offset vertically for clarity. The pattern collected at 155~K with the dashed black curve corresponds to spatially segregated regions of rhombohedral and monoclinic symmetry. }
	\end{figure*}
	To investigate the possibility of any local structural distortions persisting above the long-range structural phase transition, we inspect the neutron PDF patterns collected on a fine temperature grid across the transition. The PDFs collected between 120 and 200~K are displayed in Fig.~\ref{fig:neutronwaterfall}, offset vertically with the temperature increasing in the vertically upward direction. Panel (a) shows a wide view of the data (up to $\sim$18~\AA), while panel (b) zooms in on just the first 6~\AA. The wide view reveals an abrupt change in the characteristic shape of the PDF pattern occurring between the 150~K and 160~K, reflecting the significant rearrangement of the atomic positions at the structural phase transition. Fig.~\ref{fig:neutronwaterfall}(b) shows that this abrupt change in the observed PDF is likewise observed in the low-$r$ region of the data, which corresponds to the correlations between the first few nearest neighbors for any given atom. In particular, the prominent peak centered around 2.9~\AA\ shifts abruptly to slightly lower $r$ when the temperature is raised from 155~K to 160~K, and the peak centered around 4~\AA\ is much broader at 155~K and below, indicating the proliferation of atom pair separation distances due to symmetry breaking in the low-temperature phase. A similar effect is seen around 5.5~\AA. The data set at 155~K shows distinct features that are present at both higher and lower temperatures, again indicating that the sample consists of phase-separated rhombohedral and monoclinic regions at this temperature. These observations point to an abrupt, first-order transition from rhombohedral to monoclinic crystallographic symmetry that occurs uniformly on both local length scales and longer-range length scales; no observable distortions of the oxygen positions characteristic of the low-temperature monoclinic phase persist into the high-temperature rhombohedral phase.

	\subsection{X-ray PDF analysis}
\begin{figure*}
		\includegraphics[width=130mm]{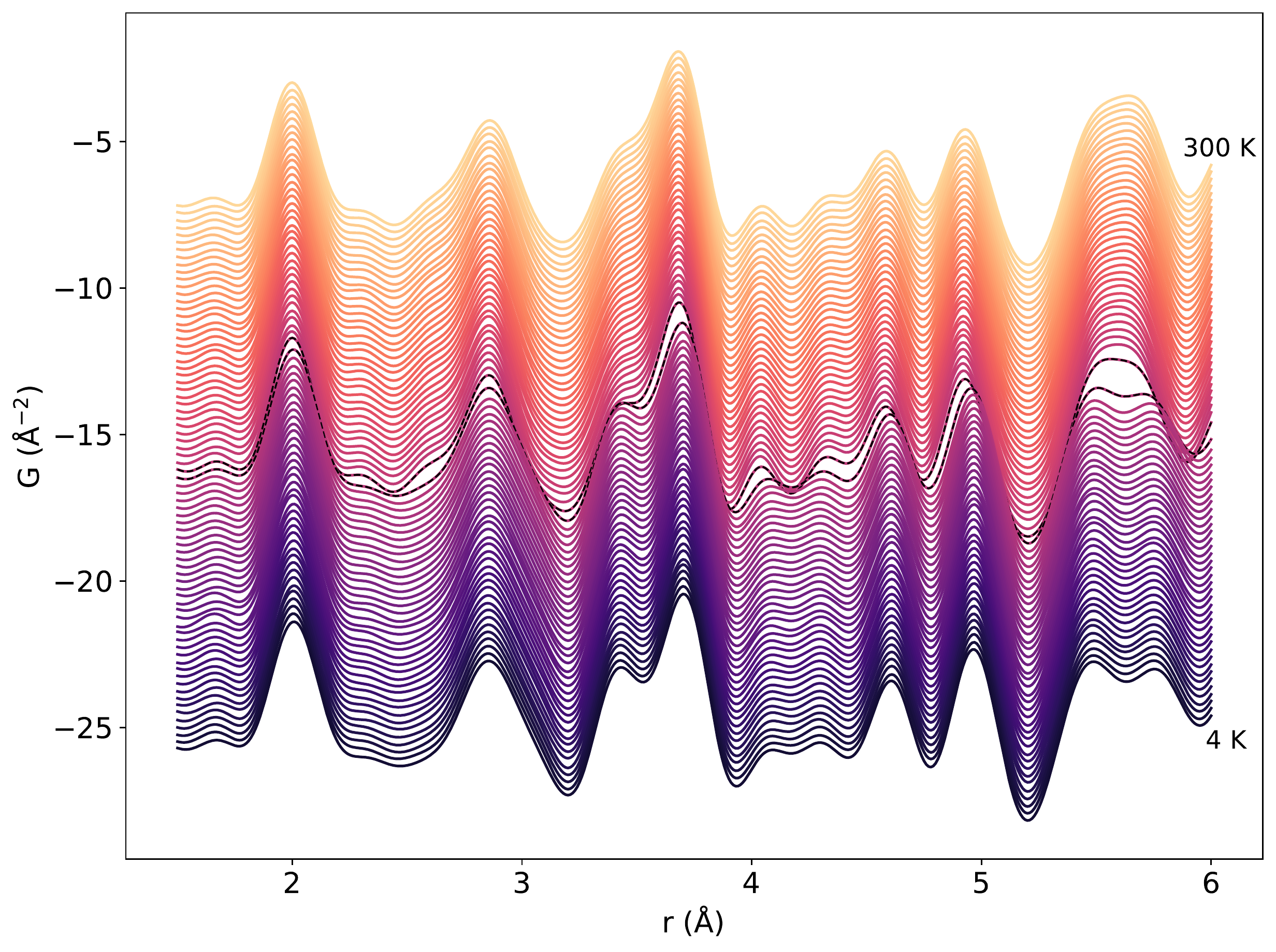}
		\caption{\label{fig:xraywaterfall} X-ray PDF patterns for \vo\ collected between 4~K and 300~K in steps of 4~K, offset vertically for clarity. The patterns change abruptly between 148~K and 160~K, with the patterns collected at 152~K and 156~K (highlighted by the black dashed curves) showing a mixture of high- and low-temperature features due to phase separation in the sample.}
	\end{figure*}

	The x-ray PDF analysis corroborates the abruptness of the structural transition and the lack of any local distortions surviving into the rhombohedral phase. In Fig.~\ref{fig:xraywaterfall} we display the x-ray PDF data between 4~K and 300~K, again offset vertically.
	As with the neutron data, sudden and significant changes occur as the temperature is raised above $\sim$150~K. The most obvious change is the low-temperature splitting of the broad feature centered on $\sim$5.6~\AA. Additional changes can be identified for the peaks centered around 4.6 and 5.0~\AA\ (peak positions shift to lower $r$ at high temperature); the small peaks between 4.1 and 4.4~\AA\ (broadening and shifting of peak centers at low temperature); the two-peak feature between 3.2 and 3.9~\AA\ (enhanced splitting at low temperature); and the peak centered on 2.8~\AA\ (broader at low temperature). Consistent with the neutron PDF, these observations indicate that no measurable distortions persist into the high temperature phase. The two patterns highlighted by the black dashed curves, which were collected at 152~K and 156~K, show features characteristic of both the low- and high-temperature structure, once again indicating coexisting rhombohedral and monoclinic regions that are spatially segregated.
	
\begin{figure}
		\includegraphics[width=80mm]{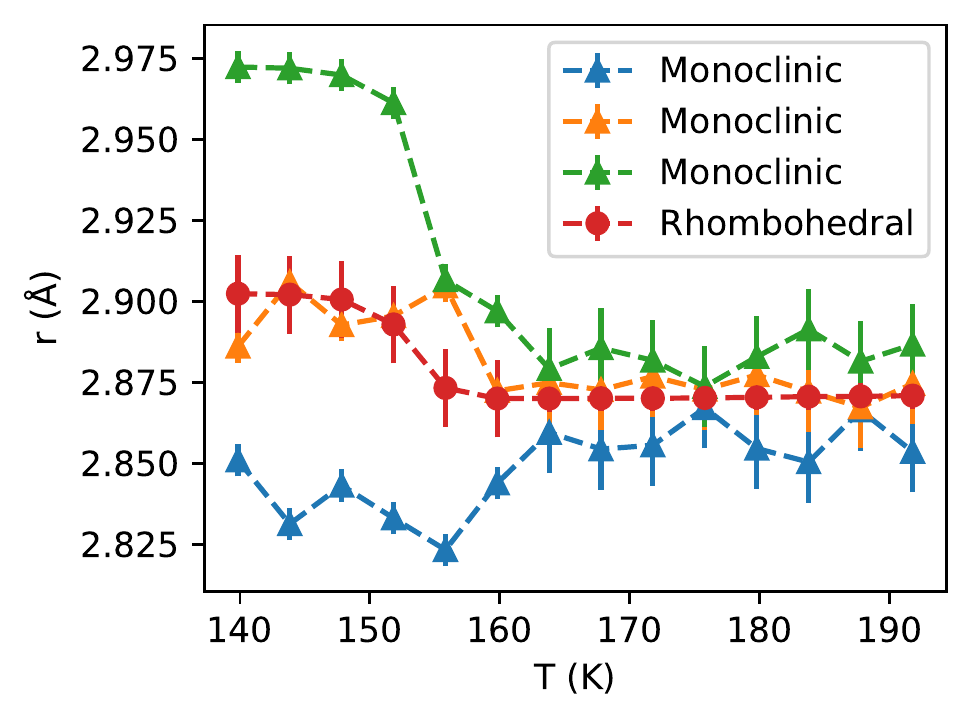}
		\caption{\label{fig:bondlength} Vanadium-vanadium pair distances extracted from fits to the first 10~\AA\ of the x-ray PDF data using the monoclinic model (triangles) and the rhombohedral model (circles). Above the transition, the monoclinic model converges to be practically indistinguishable from the rhombohedral model, confirming the lack of any persistent monoclinic distortions in the local structure. The error bars were determined from the estimated standard deviation of the fit parameters.}
		
	\end{figure}

	To provide further confirmation of this, we fit the monoclinic and rhombohedral structural models over the first 10~\AA\ of the x-ray PDF data at several temperatures both below and above the transition. From the refined structural models, we then extracted the V-V atom pair distances. The structural transition in the average structure is known to result in distinct V-V distances in the monoclinic and rhombohedral phases. If no monoclinic distortions persist into the high-temperature phase, as the results thus far have suggested, then the monoclinic model should naturally refine to be equivalent to the rhombohedral model at high temperature. In other words, the monoclinic model V-V distances should converge to those of the rhombohedral model above the transition. This is precisely what we observe, as seen in Fig.~\ref{fig:bondlength} for the second nearest neighbor V-V pair in the rhombohedral structure (orange lines in Fig.~\ref{fig:structure}). Below $\sim$160~K, the monoclinic model yields three distinct V-V distances around 2.84, 2.90, and 2.97~\AA, whereas the symmetry of the rhombohedral model only allows a single V-V distance. Above $\sim$160~K, the monoclinic V-V distances collapse onto the rhombohedral V-V distance within experimental sensitivity, once again demonstrating the lack of any persistent monoclinic distortions of the local structure above the transition. We note that the nearest neighbor V-V distance is not split into multiple distances by the monoclinic distortion, so the second nearest neighbor is the closest V-V pair with sensitivity to monoclinic symmetry breaking.
	
	\subsection{Magnetic PDF Analysis}
	The magnetic correlations in \vo\ give rise to additional intensity in the neutron scattering pattern, which likewise contributes to the real-space signal as the magnetic PDF. For completeness, we also include the magnetic PDF analysis here. The magnetic PDF was isolated by first performing atomic PDF fits at all temperatures, and then subtracting the fit residual at 300~K (where correlated magnetic scattering is negligible) from the fit residuals at all lower temperatures. This approach, which has been used successfully elsewhere~\cite{frand;prm17}, allows us to isolate the temperature-dependent portion of the total PDF signal that cannot be captured by the atomic PDF fit. In an experiment with a temperature-independent background signal, the remaining temperature-dependent signal should be dominated by the magnetic PDF (assuming the atomic PDF fits are accurate, which they are in this case). We further cleaned the magnetic PDF signal by filtering out all frequencies above 5~\AA$^{-1}$, above which the magnetic scattering is very weak due to the magnetic form factor, and by applying a dilation transformation to the low-temperature atomic PDF fit residuals to minimize the artifacts introduced by thermal expansion when subtracting the 300~K reference fit residual. The latter correction was implemented by performing a least-squares minimization of the difference between the 300~K fit residual and the low-temperature fit residual, where the dilation was applied to the low-temperature fit residual to stretch the signal in the $r$-dimension, mimicking the effect of thermal expansion. This optimized magnetic PDF signal was then used as input for the magnetic PDF fits. The published antiferromagnetic structure~\cite{moon;prl70} was used for the fits, with the only free parameter being a scale factor.
	
\begin{figure}
		\includegraphics[width=70mm]{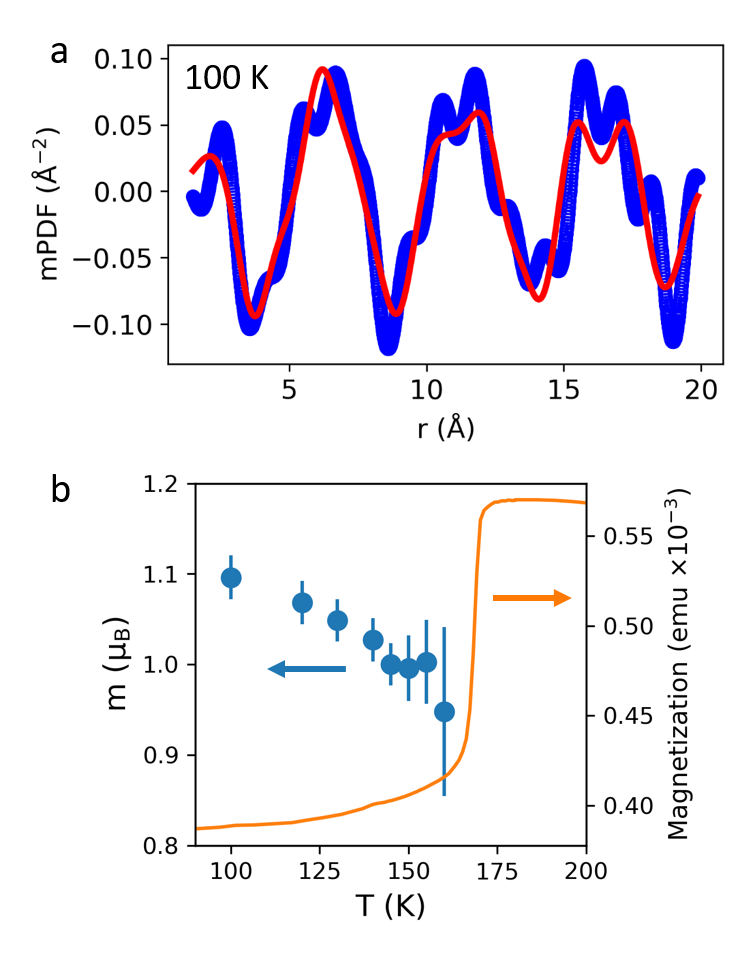}
		\caption{\label{fig:mPDF} (a) Magnetic PDF fit for \vo\ at 100~K using the published antiferromagnetic structure. The blue curve represents the experimental magnetic PDF curve isolated from the neutron PDF data as described in the text, and the red curve represents the calculated magnetic PDF. (b) The ordered magnetic moment as a function of temperature determined from the magnetic PDF fits.}
	\end{figure}

	A representative magnetic PDF fit performed at 100~K is displayed in Fig.~\ref{fig:mPDF}(a). 
	The fit is good, considering how weak the magnetic PDF signal is (over 30 times smaller in magnitude than the atomic PDF). The magnetic PDF scale factor is related to the ordered magnetic moment by~\cite{frand;aca15,kodam;jpsj17,zhang;prb19}
	\begin{align}\label{m}
		m = g\sqrt{\frac{C_{\mathrm{mag}}\langle b \rangle^2}{C_{\mathrm{at}}n_{\mathrm{mag}}}},
	\end{align}
	where $m$ is the locally ordered magnetic moment in Bohr magnetons, $g$ is the Land\'e $g$-factor (taken to be 2 in this case), $C_{\mathrm{mag}}$ and $C_{\mathrm{at}}$ are the best-fit scale factors for the magnetic and atomic PDF components, respectively, $\langle b \rangle$ is the average coherent neutron scattering length of the sample, and $n_{\mathrm{mag}}$ is the fraction of atoms in the sample with a nonzero magnetic moment. We note that this formula assumes that the magnetic moment vectors when calculating the magnetic PDF have unit length. Using this conversion factor, we display the temperature evolution of the ordered moment in Fig.~\ref{fig:mPDF}(b), revealing an abrupt, first-order transition around 160~K, as expected. Previous experiments using various magnetic probes have indicated that short-range antiferromagnetic correlations exist in the high temperature phase~\cite{bao;prl97,bao;prb98,frand;prb19b,trast;prb20}, which would in principle yield a magnetic PDF signal above the transition. However, we were unable to detect a magnetic signal above the transition, suggesting it is too weak to be seen with the current experimental sensitivity. This is unsurprising, given how weak the mPDF signal is even in the ordered state.
	
	\section{Discussion and Conclusion}
	The PDF data demonstrate that the structural transition in \vo\ occurs abruptly and uniformly on short and long length scales, without any local monoclinic distortions persisting into the rhombohedral phase. The lack of monoclinic fluctuations in the rhombohedral phase is in marked contrast to the well-documented observation of antiferromagnetic fluctuations in the paramagnetic phase~\cite{bao;prl97,bao;prb98,frand;prb19b,trast;prb20}. This suggests that a continuous magnetic transition is proximate in parameter space but is preempted by the first-order structural phase transition. Such a scenario is further supported by the magnetic PDF result showing a reduction of the ordered moment with increasing temperature in the insulating state (the expected trend for a continuous transition) until a discontinuous reduction of the ordered moment to zero occurs at the structural phase transition. Thus, the spin \dof\ is active above the transition, in contrast to the lattice \dof. These observations indicate that the spin channel is likely to be more relevant than the lattice channel for the physics of the MIT in \vo, consistent with recent theoretical and experimental work~\cite{trast;prb20}. In this scenario, the structural transition would result from the modifications of the atomic binding forces caused by the localization of the electron wave functions in the low-temperature insulating phase ~\cite{trast;prb20}. The Mott localized state leads to an expansion of the lattice (realized by the monoclinic distortion), since this decreases the electronic overlap integrals and thereby lowers the Coulomb repulsion. This is also observed in the paramagnetic metal to paramagnetic insulator phase transition of Cr-doped \vo, in which the crystallographic symmetry remains unchanged, but the lattice expands abruptly in a first-order transition~\cite{mcwha;prl69}.
	
	
	We note that the lack of short-range monoclinic distortions in the rhombohedral phase reported here differs from two previous EXAFS studies of \vo\ that reported monoclinic distortions of the local structure above the MIT~\cite{frenk;ssc97,pfalz;prb06}. We comment on possible explanations for this discrepancy. First, the reported observation of short-range monoclinic distortions in Ref.~\onlinecite{pfalz;prb06} occurs inside a narrow temperature window extending $\sim$10~K above the long-range transition. Given the observation of phase-separated regions of rhombohedral and monoclinic crystal structures in a finite temperature window as established in this study and others~\cite{kalch;prl19,frand;prb19b}, it seems plausible that this earlier EXAFS study was simply detecting this narrow temperature window of coexisting phases. Second, sample-dependent variations of the stoichiometry could also be responsible for the discrepancy. The sample of \vo\ used in Ref.~\onlinecite{pfalz;prb06} was known to be deficient in vanadium, resulting in a reduction of the MIT temperature by $\sim$15~K compared to ideal samples, whereas the sample in the current study shows no such reduction in transition temperature. The structural defects introduced by vanadium vacancies could potentially stabilize regions of local monoclinic symmetry above the transition. Further studies of this possibility would be interesting to pursue. This may also be the case in Ref.~\onlinecite{frenk;ssc97}, where it is reported that attempts were made to improve the stoichiometry during the synthesis process, but further details were not provided.
	
	

	\textbf{Acknowledgements}
	
	We thank Milinda Abeykoon for his assistance with x-ray PDF experiments at NSLS-II, along with Michelle Everett, J\"org Neuefeind, and Jue Liu for their assistance with neutron PDF experiments at the SNS. E.R.A.F. was supported by the College of Physical and Mathematical Sciences at Brigham Young University. The magnetic pair distribution function work was supported by the U.S. Department of Energy, Office of Science, Basic Energy Science through Award No. DE-SC0021134. Y.K. acknowledges funding from the Norman
	Seiden Fellowship for Nanotechnology and Optoelectronics and the ISRAEL SCIENCE FOUNDATION (Grant No. 1031/21). This study used resources at the Spallation Neutron Source (SNS), a DOE Office of Science User Facility operated by the Oak Ridge National Laboratory. This research used beamline 28-ID-1 of the National Synchrotron Light Source II, a U.S. Department of Energy (DOE) Office of Science User Facility operated for the DOE Office of Science by Brookhaven National Laboratory under Contract No. DE-SC0012704.


\begin{thebibliography}{65}%
	\makeatletter
	\providecommand \@ifxundefined [1]{%
		\@ifx{#1\undefined}
	}%
	\providecommand \@ifnum [1]{%
		\ifnum #1\expandafter \@firstoftwo
		\else \expandafter \@secondoftwo
		\fi
	}%
	\providecommand \@ifx [1]{%
		\ifx #1\expandafter \@firstoftwo
		\else \expandafter \@secondoftwo
		\fi
	}%
	\providecommand \natexlab [1]{#1}%
	\providecommand \enquote  [1]{``#1''}%
	\providecommand \bibnamefont  [1]{#1}%
	\providecommand \bibfnamefont [1]{#1}%
	\providecommand \citenamefont [1]{#1}%
	\providecommand \href@noop [0]{\@secondoftwo}%
	\providecommand \href [0]{\begingroup \@sanitize@url \@href}%
	\providecommand \@href[1]{\@@startlink{#1}\@@href}%
	\providecommand \@@href[1]{\endgroup#1\@@endlink}%
	\providecommand \@sanitize@url [0]{\catcode `\\12\catcode `\$12\catcode
		`\&12\catcode `\#12\catcode `\^12\catcode `\_12\catcode `\%12\relax}%
	\providecommand \@@startlink[1]{}%
	\providecommand \@@endlink[0]{}%
	\providecommand \url  [0]{\begingroup\@sanitize@url \@url }%
	\providecommand \@url [1]{\endgroup\@href {#1}{\urlprefix }}%
	\providecommand \urlprefix  [0]{URL }%
	\providecommand \Eprint [0]{\href }%
	\providecommand \doibase [0]{http://dx.doi.org/}%
	\providecommand \selectlanguage [0]{\@gobble}%
	\providecommand \bibinfo  [0]{\@secondoftwo}%
	\providecommand \bibfield  [0]{\@secondoftwo}%
	\providecommand \translation [1]{[#1]}%
	\providecommand \BibitemOpen [0]{}%
	\providecommand \bibitemStop [0]{}%
	\providecommand \bibitemNoStop [0]{.\EOS\space}%
	\providecommand \EOS [0]{\spacefactor3000\relax}%
	\providecommand \BibitemShut  [1]{\csname bibitem#1\endcsname}%
	\let\auto@bib@innerbib\@empty
	\bibitem [{\citenamefont {Imada}\ \emph {et~al.}(1998)\citenamefont {Imada},
		\citenamefont {Fujimori},\ and\ \citenamefont {Tokura}}]{imada;rmp98}%
	\BibitemOpen
	\bibfield  {author} {\bibinfo {author} {\bibfnamefont {M.}~\bibnamefont
			{Imada}}, \bibinfo {author} {\bibfnamefont {A.}~\bibnamefont {Fujimori}}, \
		and\ \bibinfo {author} {\bibfnamefont {Y.}~\bibnamefont {Tokura}},\ }\href
	{\doibase 10.1103/RevModPhys.70.1039} {\bibfield  {journal} {\bibinfo
			{journal} {Rev. Mod. Phys.}\ }\textbf {\bibinfo {volume} {70}},\ \bibinfo
		{pages} {1039} (\bibinfo {year} {1998})}\BibitemShut {NoStop}%
	\bibitem [{\citenamefont {Keimer}\ and\ \citenamefont
		{Moore}(2017)}]{keime;np17}%
	\BibitemOpen
	\bibfield  {author} {\bibinfo {author} {\bibfnamefont {B.}~\bibnamefont
			{Keimer}}\ and\ \bibinfo {author} {\bibfnamefont {J.}~\bibnamefont {Moore}},\
	}\href@noop {} {\bibfield  {journal} {\bibinfo  {journal} {Nature Phys.}\
		}\textbf {\bibinfo {volume} {13}},\ \bibinfo {pages} {1045} (\bibinfo {year}
		{2017})}\BibitemShut {NoStop}%
	\bibitem [{\citenamefont {Tokura}(2000)}]{tokur;s00}%
	\BibitemOpen
	\bibfield  {author} {\bibinfo {author} {\bibfnamefont {N.}~\bibnamefont
			{Tokura}, \bibfnamefont {Y.~Nagaosa}},\ }\href@noop {} {\bibfield  {journal}
		{\bibinfo  {journal} {Science}\ }\textbf {\bibinfo {volume} {288}},\ \bibinfo
		{pages} {462} (\bibinfo {year} {2000})}\BibitemShut {NoStop}%
	\bibitem [{\citenamefont {McWhan}\ \emph {et~al.}(1969)\citenamefont {McWhan},
		\citenamefont {Rice},\ and\ \citenamefont {Remeika}}]{mcwha;prl69}%
	\BibitemOpen
	\bibfield  {author} {\bibinfo {author} {\bibfnamefont {D.~B.}\ \bibnamefont
			{McWhan}}, \bibinfo {author} {\bibfnamefont {T.~M.}\ \bibnamefont {Rice}}, \
		and\ \bibinfo {author} {\bibfnamefont {J.~P.}\ \bibnamefont {Remeika}},\
	}\href {\doibase 10.1103/PhysRevLett.23.1384} {\bibfield  {journal} {\bibinfo
			{journal} {Phys. Rev. Lett.}\ }\textbf {\bibinfo {volume} {23}},\ \bibinfo
		{pages} {1384} (\bibinfo {year} {1969})}\BibitemShut {NoStop}%
	\bibitem [{\citenamefont {McWhan}\ \emph {et~al.}(1971)\citenamefont {McWhan},
		\citenamefont {Remeika}, \citenamefont {Rice}, \citenamefont {Brinkman},
		\citenamefont {Maita},\ and\ \citenamefont {Menth}}]{mcwha;prl71}%
	\BibitemOpen
	\bibfield  {author} {\bibinfo {author} {\bibfnamefont {D.~B.}\ \bibnamefont
			{McWhan}}, \bibinfo {author} {\bibfnamefont {J.~P.}\ \bibnamefont {Remeika}},
		\bibinfo {author} {\bibfnamefont {T.~M.}\ \bibnamefont {Rice}}, \bibinfo
		{author} {\bibfnamefont {W.~F.}\ \bibnamefont {Brinkman}}, \bibinfo {author}
		{\bibfnamefont {J.~P.}\ \bibnamefont {Maita}}, \ and\ \bibinfo {author}
		{\bibfnamefont {A.}~\bibnamefont {Menth}},\ }\href {\doibase
		10.1103/PhysRevLett.27.941} {\bibfield  {journal} {\bibinfo  {journal} {Phys.
				Rev. Lett.}\ }\textbf {\bibinfo {volume} {27}},\ \bibinfo {pages} {941}
		(\bibinfo {year} {1971})}\BibitemShut {NoStop}%
	\bibitem [{\citenamefont {McWhan}\ \emph {et~al.}(9733)\citenamefont {McWhan},
		\citenamefont {Menth}, \citenamefont {Remeika}, \citenamefont {Brinkman},\
		and\ \citenamefont {Rice}}]{mcwha;prb73}%
	\BibitemOpen
	\bibfield  {author} {\bibinfo {author} {\bibfnamefont {D.~B.}\ \bibnamefont
			{McWhan}}, \bibinfo {author} {\bibfnamefont {A.}~\bibnamefont {Menth}},
		\bibinfo {author} {\bibfnamefont {J.~P.}\ \bibnamefont {Remeika}}, \bibinfo
		{author} {\bibfnamefont {W.~F.}\ \bibnamefont {Brinkman}}, \ and\ \bibinfo
		{author} {\bibfnamefont {T.~M.}\ \bibnamefont {Rice}},\ }\href@noop {}
	{\bibfield  {journal} {\bibinfo  {journal} {Phys. Rev. B}\ }\textbf {\bibinfo
			{volume} {7}},\ \bibinfo {pages} {1920 } (\bibinfo {year}
		{19733})}\BibitemShut {NoStop}%
	\bibitem [{\citenamefont {Dernier}(1970)}]{derni;jpcs70}%
	\BibitemOpen
	\bibfield  {author} {\bibinfo {author} {\bibfnamefont {P.~D.}\ \bibnamefont
			{Dernier}},\ }\href@noop {} {\bibfield  {journal} {\bibinfo  {journal} {J.
				Phys. Chem. Solids}\ }\textbf {\bibinfo {volume} {31}},\ \bibinfo {pages}
		{2569} (\bibinfo {year} {1970})}\BibitemShut {NoStop}%
	\bibitem [{\citenamefont {Moon}(1970)}]{moon;prl70}%
	\BibitemOpen
	\bibfield  {author} {\bibinfo {author} {\bibfnamefont {R.~M.}\ \bibnamefont
			{Moon}},\ }\href {\doibase 10.1103/PhysRevLett.25.527} {\bibfield  {journal}
		{\bibinfo  {journal} {Phys. Rev. Lett.}\ }\textbf {\bibinfo {volume} {25}},\
		\bibinfo {pages} {527} (\bibinfo {year} {1970})}\BibitemShut {NoStop}%
	\bibitem [{\citenamefont {Uemura}\ \emph {et~al.}(1984)\citenamefont {Uemura},
		\citenamefont {Yamazaki}, \citenamefont {Kitaoka}, \citenamefont {Takigawa},\
		and\ \citenamefont {Yasuoka}}]{uemur;hfi84}%
	\BibitemOpen
	\bibfield  {author} {\bibinfo {author} {\bibfnamefont {Y.~J.}\ \bibnamefont
			{Uemura}}, \bibinfo {author} {\bibfnamefont {T.}~\bibnamefont {Yamazaki}},
		\bibinfo {author} {\bibfnamefont {Y.}~\bibnamefont {Kitaoka}}, \bibinfo
		{author} {\bibfnamefont {M.}~\bibnamefont {Takigawa}}, \ and\ \bibinfo
		{author} {\bibfnamefont {H.}~\bibnamefont {Yasuoka}},\ }\href@noop {}
	{\bibfield  {journal} {\bibinfo  {journal} {Hyperfine Interact.}\ }\textbf
		{\bibinfo {volume} {17}},\ \bibinfo {pages} {339} (\bibinfo {year}
		{1984})}\BibitemShut {NoStop}%
	\bibitem [{\citenamefont {Denison}\ \emph {et~al.}(1985)\citenamefont
		{Denison}, \citenamefont {Boekema}, \citenamefont {Lichti}, \citenamefont
		{Chan}, \citenamefont {Cooke}, \citenamefont {Heffner}, \citenamefont
		{Hutson}, \citenamefont {Leon},\ and\ \citenamefont
		{Schillaci}}]{denis;jap85}%
	\BibitemOpen
	\bibfield  {author} {\bibinfo {author} {\bibfnamefont {A.~B.}\ \bibnamefont
			{Denison}}, \bibinfo {author} {\bibfnamefont {C.}~\bibnamefont {Boekema}},
		\bibinfo {author} {\bibfnamefont {R.~L.}\ \bibnamefont {Lichti}}, \bibinfo
		{author} {\bibfnamefont {K.~C.}\ \bibnamefont {Chan}}, \bibinfo {author}
		{\bibfnamefont {D.~W.}\ \bibnamefont {Cooke}}, \bibinfo {author}
		{\bibfnamefont {R.~H.}\ \bibnamefont {Heffner}}, \bibinfo {author}
		{\bibfnamefont {R.~L.}\ \bibnamefont {Hutson}}, \bibinfo {author}
		{\bibfnamefont {M.}~\bibnamefont {Leon}}, \ and\ \bibinfo {author}
		{\bibfnamefont {M.~E.}\ \bibnamefont {Schillaci}},\ }\href {\doibase
		10.1063/1.334955} {\bibfield  {journal} {\bibinfo  {journal} {J. Appl.
				Phys.}\ }\textbf {\bibinfo {volume} {57}},\ \bibinfo {pages} {3743} (\bibinfo
		{year} {1985})},\ \Eprint
	{http://arxiv.org/abs/https://doi.org/10.1063/1.334955}
	{https://doi.org/10.1063/1.334955} \BibitemShut {NoStop}%
	\bibitem [{\citenamefont {Bao}\ \emph {et~al.}(1993)\citenamefont {Bao},
		\citenamefont {Broholm}, \citenamefont {Carter}, \citenamefont {Rosenbaum},
		\citenamefont {Aeppli}, \citenamefont {Trevino}, \citenamefont {Metcalf},
		\citenamefont {Honig},\ and\ \citenamefont {Spalek}}]{bao;prl93}%
	\BibitemOpen
	\bibfield  {author} {\bibinfo {author} {\bibfnamefont {W.}~\bibnamefont
			{Bao}}, \bibinfo {author} {\bibfnamefont {C.}~\bibnamefont {Broholm}},
		\bibinfo {author} {\bibfnamefont {S.~A.}\ \bibnamefont {Carter}}, \bibinfo
		{author} {\bibfnamefont {T.~F.}\ \bibnamefont {Rosenbaum}}, \bibinfo {author}
		{\bibfnamefont {G.}~\bibnamefont {Aeppli}}, \bibinfo {author} {\bibfnamefont
			{S.~F.}\ \bibnamefont {Trevino}}, \bibinfo {author} {\bibfnamefont
			{P.}~\bibnamefont {Metcalf}}, \bibinfo {author} {\bibfnamefont {J.~M.}\
			\bibnamefont {Honig}}, \ and\ \bibinfo {author} {\bibfnamefont
			{J.}~\bibnamefont {Spalek}},\ }\href {\doibase 10.1103/PhysRevLett.71.766}
	{\bibfield  {journal} {\bibinfo  {journal} {Phys. Rev. Lett.}\ }\textbf
		{\bibinfo {volume} {71}},\ \bibinfo {pages} {766} (\bibinfo {year}
		{1993})}\BibitemShut {NoStop}%
	\bibitem [{\citenamefont {Frandsen}\ \emph
		{et~al.}(2016{\natexlab{a}})\citenamefont {Frandsen}, \citenamefont {Liu},
		\citenamefont {Cheung}, \citenamefont {Guguchia}, \citenamefont {Khasanov},
		\citenamefont {Morenzoni}, \citenamefont {Munsie}, \citenamefont {Hallas},
		\citenamefont {Wilson}, \citenamefont {Cai}, \citenamefont {Luke},
		\citenamefont {Chen}, \citenamefont {Li}, \citenamefont {Jin}, \citenamefont
		{Ding}, \citenamefont {Guo}, \citenamefont {Ning}, \citenamefont {Ito},
		\citenamefont {Higemoto}, \citenamefont {Billinge}, \citenamefont {Sakamoto},
		\citenamefont {Fujimori}, \citenamefont {Murakami}, \citenamefont {Kageyama},
		\citenamefont {Alonso}, \citenamefont {Kotliar}, \citenamefont {Imada},\ and\
		\citenamefont {Uemura}}]{frand;nc16}%
	\BibitemOpen
	\bibfield  {author} {\bibinfo {author} {\bibfnamefont {B.~A.}\ \bibnamefont
			{Frandsen}}, \bibinfo {author} {\bibfnamefont {L.}~\bibnamefont {Liu}},
		\bibinfo {author} {\bibfnamefont {S.~C.}\ \bibnamefont {Cheung}}, \bibinfo
		{author} {\bibfnamefont {Z.}~\bibnamefont {Guguchia}}, \bibinfo {author}
		{\bibfnamefont {R.}~\bibnamefont {Khasanov}}, \bibinfo {author}
		{\bibfnamefont {E.}~\bibnamefont {Morenzoni}}, \bibinfo {author}
		{\bibfnamefont {T.~J.~S.}\ \bibnamefont {Munsie}}, \bibinfo {author}
		{\bibfnamefont {A.~M.}\ \bibnamefont {Hallas}}, \bibinfo {author}
		{\bibfnamefont {M.~N.}\ \bibnamefont {Wilson}}, \bibinfo {author}
		{\bibfnamefont {Y.}~\bibnamefont {Cai}}, \bibinfo {author} {\bibfnamefont
			{G.~M.}\ \bibnamefont {Luke}}, \bibinfo {author} {\bibfnamefont
			{B.}~\bibnamefont {Chen}}, \bibinfo {author} {\bibfnamefont {W.}~\bibnamefont
			{Li}}, \bibinfo {author} {\bibfnamefont {C.}~\bibnamefont {Jin}}, \bibinfo
		{author} {\bibfnamefont {C.}~\bibnamefont {Ding}}, \bibinfo {author}
		{\bibfnamefont {S.}~\bibnamefont {Guo}}, \bibinfo {author} {\bibfnamefont
			{F.}~\bibnamefont {Ning}}, \bibinfo {author} {\bibfnamefont {T.}~\bibnamefont
			{Ito}}, \bibinfo {author} {\bibfnamefont {W.}~\bibnamefont {Higemoto}},
		\bibinfo {author} {\bibfnamefont {S.~J.~L.}\ \bibnamefont {Billinge}},
		\bibinfo {author} {\bibfnamefont {S.}~\bibnamefont {Sakamoto}}, \bibinfo
		{author} {\bibfnamefont {A.}~\bibnamefont {Fujimori}}, \bibinfo {author}
		{\bibfnamefont {T.}~\bibnamefont {Murakami}}, \bibinfo {author}
		{\bibfnamefont {H.}~\bibnamefont {Kageyama}}, \bibinfo {author}
		{\bibfnamefont {J.~A.}\ \bibnamefont {Alonso}}, \bibinfo {author}
		{\bibfnamefont {G.}~\bibnamefont {Kotliar}}, \bibinfo {author} {\bibfnamefont
			{M.}~\bibnamefont {Imada}}, \ and\ \bibinfo {author} {\bibfnamefont {Y.~J.}\
			\bibnamefont {Uemura}},\ }\href {\doibase 10.1038/ncomms12519} {\bibfield
		{journal} {\bibinfo  {journal} {Nat. Commun.}\ }\textbf {\bibinfo {volume}
			{7}},\ \bibinfo {pages} {12519} (\bibinfo {year}
		{2016}{\natexlab{a}})}\BibitemShut {NoStop}%
	\bibitem [{\citenamefont {Gossard}\ \emph {et~al.}(1970)\citenamefont
		{Gossard}, \citenamefont {McWhan},\ and\ \citenamefont
		{Remeika}}]{gossa;prb70}%
	\BibitemOpen
	\bibfield  {author} {\bibinfo {author} {\bibfnamefont {A.~C.}\ \bibnamefont
			{Gossard}}, \bibinfo {author} {\bibfnamefont {D.~B.}\ \bibnamefont {McWhan}},
		\ and\ \bibinfo {author} {\bibfnamefont {J.~P.}\ \bibnamefont {Remeika}},\
	}\href@noop {} {\bibfield  {journal} {\bibinfo  {journal} {Phys. Rev. B}\
		}\textbf {\bibinfo {volume} {2}},\ \bibinfo {pages} {3762} (\bibinfo {year}
		{1970})}\BibitemShut {NoStop}%
	\bibitem [{\citenamefont {Ueda}\ \emph {et~al.}(1978)\citenamefont {Ueda},
		\citenamefont {Kosuge}, \citenamefont {Kachi}, \citenamefont {Nishihara},\
		and\ \citenamefont {Heidemann}}]{ueda;jpcs78}%
	\BibitemOpen
	\bibfield  {author} {\bibinfo {author} {\bibfnamefont {Y.}~\bibnamefont
			{Ueda}}, \bibinfo {author} {\bibfnamefont {K.}~\bibnamefont {Kosuge}},
		\bibinfo {author} {\bibfnamefont {H.}~\bibnamefont {Kachi}, \bibfnamefont
			{S.~Yasuoka}}, \bibinfo {author} {\bibfnamefont {H.}~\bibnamefont
			{Nishihara}}, \ and\ \bibinfo {author} {\bibfnamefont {A.}~\bibnamefont
			{Heidemann}},\ }\href@noop {} {\bibfield  {journal} {\bibinfo  {journal} {J.
				Phys. Chem. Solids}\ }\textbf {\bibinfo {volume} {39}},\ \bibinfo {pages}
		{1281} (\bibinfo {year} {1978})}\BibitemShut {NoStop}%
	\bibitem [{\citenamefont {Grygiel}\ \emph {et~al.}(2007)\citenamefont
		{Grygiel}, \citenamefont {Simon}, \citenamefont {Mercey}, \citenamefont
		{Prellier},\ and\ \citenamefont {Fr\'esard}}]{grygi;apl07}%
	\BibitemOpen
	\bibfield  {author} {\bibinfo {author} {\bibfnamefont {C.}~\bibnamefont
			{Grygiel}}, \bibinfo {author} {\bibfnamefont {C.}~\bibnamefont {Simon}},
		\bibinfo {author} {\bibfnamefont {B.}~\bibnamefont {Mercey}}, \bibinfo
		{author} {\bibfnamefont {W.}~\bibnamefont {Prellier}}, \ and\ \bibinfo
		{author} {\bibfnamefont {R.}~\bibnamefont {Fr\'esard}},\ }\href@noop {}
	{\bibfield  {journal} {\bibinfo  {journal} {Appl. Phys. Lett.}\ }\textbf
		{\bibinfo {volume} {91}},\ \bibinfo {pages} {262103} (\bibinfo {year}
		{2007})}\BibitemShut {NoStop}%
	\bibitem [{\citenamefont {Ji}\ \emph {et~al.}(2012)\citenamefont {Ji},
		\citenamefont {Pan}, \citenamefont {Bi}, \citenamefont {Liang}, \citenamefont
		{Zhang}, \citenamefont {Zeng}, \citenamefont {Wen}, \citenamefont {Zhang},
		\citenamefont {Chen}, \citenamefont {Jia},\ and\ \citenamefont
		{Lin}}]{ji;apl12}%
	\BibitemOpen
	\bibfield  {author} {\bibinfo {author} {\bibfnamefont {Y.~D.}\ \bibnamefont
			{Ji}}, \bibinfo {author} {\bibfnamefont {T.~S.}\ \bibnamefont {Pan}},
		\bibinfo {author} {\bibfnamefont {Z.}~\bibnamefont {Bi}}, \bibinfo {author}
		{\bibfnamefont {W.~Z.}\ \bibnamefont {Liang}}, \bibinfo {author}
		{\bibfnamefont {Y.}~\bibnamefont {Zhang}}, \bibinfo {author} {\bibfnamefont
			{H.~Z.}\ \bibnamefont {Zeng}}, \bibinfo {author} {\bibfnamefont {Q.~Y.}\
			\bibnamefont {Wen}}, \bibinfo {author} {\bibfnamefont {H.~W.}\ \bibnamefont
			{Zhang}}, \bibinfo {author} {\bibfnamefont {C.~L.}\ \bibnamefont {Chen}},
		\bibinfo {author} {\bibfnamefont {Q.~X.}\ \bibnamefont {Jia}}, \ and\
		\bibinfo {author} {\bibfnamefont {Y.}~\bibnamefont {Lin}},\ }\href@noop {}
	{\bibfield  {journal} {\bibinfo  {journal} {Appl. Phys. Lett.}\ }\textbf
		{\bibinfo {volume} {101}},\ \bibinfo {pages} {071902} (\bibinfo {year}
		{2012})}\BibitemShut {NoStop}%
	\bibitem [{\citenamefont {Dillemans}\ \emph {et~al.}(2014)\citenamefont
		{Dillemans}, \citenamefont {Smets}, \citenamefont {Lieten}, \citenamefont
		{Menghini}, \citenamefont {Su},\ and\ \citenamefont {Locquet}}]{dille;apl14}%
	\BibitemOpen
	\bibfield  {author} {\bibinfo {author} {\bibfnamefont {L.}~\bibnamefont
			{Dillemans}}, \bibinfo {author} {\bibfnamefont {T.}~\bibnamefont {Smets}},
		\bibinfo {author} {\bibfnamefont {R.~R.}\ \bibnamefont {Lieten}}, \bibinfo
		{author} {\bibfnamefont {M.}~\bibnamefont {Menghini}}, \bibinfo {author}
		{\bibfnamefont {C.-Y.}\ \bibnamefont {Su}}, \ and\ \bibinfo {author}
		{\bibfnamefont {J.-P.}\ \bibnamefont {Locquet}},\ }\href@noop {} {\bibfield
		{journal} {\bibinfo  {journal} {Appl. Phys. Lett.}\ }\textbf {\bibinfo
			{volume} {104}},\ \bibinfo {pages} {071902} (\bibinfo {year}
		{2014})}\BibitemShut {NoStop}%
	\bibitem [{\citenamefont {Kalcheim}\ \emph {et~al.}(2020)\citenamefont
		{Kalcheim}, \citenamefont {Adda}, \citenamefont {Salev}, \citenamefont {Lee},
		\citenamefont {Ghazikhanian}, \citenamefont {Vargas}, \citenamefont {del
			Valle},\ and\ \citenamefont {Schuller}}]{kalch;afm20}%
	\BibitemOpen
	\bibfield  {author} {\bibinfo {author} {\bibfnamefont {Y.}~\bibnamefont
			{Kalcheim}}, \bibinfo {author} {\bibfnamefont {C.}~\bibnamefont {Adda}},
		\bibinfo {author} {\bibfnamefont {P.}~\bibnamefont {Salev}}, \bibinfo
		{author} {\bibfnamefont {M.-H.}\ \bibnamefont {Lee}}, \bibinfo {author}
		{\bibfnamefont {N.}~\bibnamefont {Ghazikhanian}}, \bibinfo {author}
		{\bibfnamefont {N.~M.}\ \bibnamefont {Vargas}}, \bibinfo {author}
		{\bibfnamefont {J.}~\bibnamefont {del Valle}}, \ and\ \bibinfo {author}
		{\bibfnamefont {I.~K.}\ \bibnamefont {Schuller}},\ }\href {\doibase
		https://doi.org/10.1002/adfm.202005939} {\bibfield  {journal} {\bibinfo
			{journal} {Adv. Funct. Mater.}\ }\textbf {\bibinfo {volume} {30}},\ \bibinfo
		{pages} {2005939} (\bibinfo {year} {2020})}\BibitemShut {NoStop}%
	\bibitem [{\citenamefont {Lee}\ \emph {et~al.}(2021)\citenamefont {Lee},
		\citenamefont {Kalcheim}, \citenamefont {del Valle},\ and\ \citenamefont
		{Schuller}}]{lee;acsami21}%
	\BibitemOpen
	\bibfield  {author} {\bibinfo {author} {\bibfnamefont {M.-H.}\ \bibnamefont
			{Lee}}, \bibinfo {author} {\bibfnamefont {Y.}~\bibnamefont {Kalcheim}},
		\bibinfo {author} {\bibfnamefont {J.}~\bibnamefont {del Valle}}, \ and\
		\bibinfo {author} {\bibfnamefont {I.~K.}\ \bibnamefont {Schuller}},\ }\href
	{\doibase 10.1021/acsami.0c18327} {\bibfield  {journal} {\bibinfo  {journal}
			{ACS Appl. Mater. Interfaces}\ }\textbf {\bibinfo {volume} {13}},\ \bibinfo
		{pages} {887} (\bibinfo {year} {2021})}\BibitemShut {NoStop}%
	\bibitem [{\citenamefont {Castellani}\ \emph
		{et~al.}(1978{\natexlab{a}})\citenamefont {Castellani}, \citenamefont
		{Natoli},\ and\ \citenamefont {Ranninger}}]{caste;prb78a}%
	\BibitemOpen
	\bibfield  {author} {\bibinfo {author} {\bibfnamefont {C.}~\bibnamefont
			{Castellani}}, \bibinfo {author} {\bibfnamefont {C.~R.}\ \bibnamefont
			{Natoli}}, \ and\ \bibinfo {author} {\bibfnamefont {J.}~\bibnamefont
			{Ranninger}},\ }\href {\doibase 10.1103/PhysRevB.18.4945} {\bibfield
		{journal} {\bibinfo  {journal} {Phys. Rev. B}\ }\textbf {\bibinfo {volume}
			{18}},\ \bibinfo {pages} {4945} (\bibinfo {year}
		{1978}{\natexlab{a}})}\BibitemShut {NoStop}%
	\bibitem [{\citenamefont {Castellani}\ \emph
		{et~al.}(1978{\natexlab{b}})\citenamefont {Castellani}, \citenamefont
		{Natoli},\ and\ \citenamefont {Ranninger}}]{caste;prb78b}%
	\BibitemOpen
	\bibfield  {author} {\bibinfo {author} {\bibfnamefont {C.}~\bibnamefont
			{Castellani}}, \bibinfo {author} {\bibfnamefont {C.~R.}\ \bibnamefont
			{Natoli}}, \ and\ \bibinfo {author} {\bibfnamefont {J.}~\bibnamefont
			{Ranninger}},\ }\href {\doibase 10.1103/PhysRevB.18.4967} {\bibfield
		{journal} {\bibinfo  {journal} {Phys. Rev. B}\ }\textbf {\bibinfo {volume}
			{18}},\ \bibinfo {pages} {4967} (\bibinfo {year}
		{1978}{\natexlab{b}})}\BibitemShut {NoStop}%
	\bibitem [{\citenamefont {Castellani}\ \emph
		{et~al.}(1978{\natexlab{c}})\citenamefont {Castellani}, \citenamefont
		{Natoli},\ and\ \citenamefont {Ranninger}}]{caste;prb78c}%
	\BibitemOpen
	\bibfield  {author} {\bibinfo {author} {\bibfnamefont {C.}~\bibnamefont
			{Castellani}}, \bibinfo {author} {\bibfnamefont {C.~R.}\ \bibnamefont
			{Natoli}}, \ and\ \bibinfo {author} {\bibfnamefont {J.}~\bibnamefont
			{Ranninger}},\ }\href {\doibase 10.1103/PhysRevB.18.5001} {\bibfield
		{journal} {\bibinfo  {journal} {Phys. Rev. B}\ }\textbf {\bibinfo {volume}
			{18}},\ \bibinfo {pages} {5001} (\bibinfo {year}
		{1978}{\natexlab{c}})}\BibitemShut {NoStop}%
	\bibitem [{\citenamefont {Paolasini}\ \emph {et~al.}(1999)\citenamefont
		{Paolasini}, \citenamefont {Vettier}, \citenamefont {de~Bergevin},
		\citenamefont {Yakhou}, \citenamefont {Mannix}, \citenamefont {Stunault},
		\citenamefont {Neubeck}, \citenamefont {Altarelli}, \citenamefont {Fabrizio},
		\citenamefont {Metcalf},\ and\ \citenamefont {Honig}}]{paola;prl99}%
	\BibitemOpen
	\bibfield  {author} {\bibinfo {author} {\bibfnamefont {L.}~\bibnamefont
			{Paolasini}}, \bibinfo {author} {\bibfnamefont {C.}~\bibnamefont {Vettier}},
		\bibinfo {author} {\bibfnamefont {F.}~\bibnamefont {de~Bergevin}}, \bibinfo
		{author} {\bibfnamefont {F.}~\bibnamefont {Yakhou}}, \bibinfo {author}
		{\bibfnamefont {D.}~\bibnamefont {Mannix}}, \bibinfo {author} {\bibfnamefont
			{A.}~\bibnamefont {Stunault}}, \bibinfo {author} {\bibfnamefont
			{W.}~\bibnamefont {Neubeck}}, \bibinfo {author} {\bibfnamefont
			{M.}~\bibnamefont {Altarelli}}, \bibinfo {author} {\bibfnamefont
			{M.}~\bibnamefont {Fabrizio}}, \bibinfo {author} {\bibfnamefont {P.~A.}\
			\bibnamefont {Metcalf}}, \ and\ \bibinfo {author} {\bibfnamefont {J.~M.}\
			\bibnamefont {Honig}},\ }\href {\doibase 10.1103/PhysRevLett.82.4719}
	{\bibfield  {journal} {\bibinfo  {journal} {Phys. Rev. Lett.}\ }\textbf
		{\bibinfo {volume} {82}},\ \bibinfo {pages} {4719} (\bibinfo {year}
		{1999})}\BibitemShut {NoStop}%
	\bibitem [{\citenamefont {Park}\ \emph {et~al.}(2000)\citenamefont {Park},
		\citenamefont {Tjeng}, \citenamefont {Tanaka}, \citenamefont {Allen},
		\citenamefont {Chen}, \citenamefont {Metcalf}, \citenamefont {Honig},
		\citenamefont {de~Groot},\ and\ \citenamefont {Sawatzky}}]{park;prb00}%
	\BibitemOpen
	\bibfield  {author} {\bibinfo {author} {\bibfnamefont {J.-H.}\ \bibnamefont
			{Park}}, \bibinfo {author} {\bibfnamefont {L.~H.}\ \bibnamefont {Tjeng}},
		\bibinfo {author} {\bibfnamefont {A.}~\bibnamefont {Tanaka}}, \bibinfo
		{author} {\bibfnamefont {J.~W.}\ \bibnamefont {Allen}}, \bibinfo {author}
		{\bibfnamefont {C.~T.}\ \bibnamefont {Chen}}, \bibinfo {author}
		{\bibfnamefont {P.}~\bibnamefont {Metcalf}}, \bibinfo {author} {\bibfnamefont
			{J.~M.}\ \bibnamefont {Honig}}, \bibinfo {author} {\bibfnamefont {F.~M.~F.}\
			\bibnamefont {de~Groot}}, \ and\ \bibinfo {author} {\bibfnamefont {G.~A.}\
			\bibnamefont {Sawatzky}},\ }\href {\doibase 10.1103/PhysRevB.61.11506}
	{\bibfield  {journal} {\bibinfo  {journal} {Phys. Rev. B}\ }\textbf {\bibinfo
			{volume} {61}},\ \bibinfo {pages} {11506} (\bibinfo {year}
		{2000})}\BibitemShut {NoStop}%
	\bibitem [{\citenamefont {Rozenberg}\ \emph {et~al.}(1995)\citenamefont
		{Rozenberg}, \citenamefont {Kotliar}, \citenamefont {Kajueter}, \citenamefont
		{Thomas}, \citenamefont {Rapkine}, \citenamefont {Honig},\ and\ \citenamefont
		{Metcalf}}]{rozen;prl95}%
	\BibitemOpen
	\bibfield  {author} {\bibinfo {author} {\bibfnamefont {M.~J.}\ \bibnamefont
			{Rozenberg}}, \bibinfo {author} {\bibfnamefont {G.}~\bibnamefont {Kotliar}},
		\bibinfo {author} {\bibfnamefont {H.}~\bibnamefont {Kajueter}}, \bibinfo
		{author} {\bibfnamefont {G.~A.}\ \bibnamefont {Thomas}}, \bibinfo {author}
		{\bibfnamefont {D.~H.}\ \bibnamefont {Rapkine}}, \bibinfo {author}
		{\bibfnamefont {J.~M.}\ \bibnamefont {Honig}}, \ and\ \bibinfo {author}
		{\bibfnamefont {P.}~\bibnamefont {Metcalf}},\ }\href {\doibase
		10.1103/PhysRevLett.75.105} {\bibfield  {journal} {\bibinfo  {journal} {Phys.
				Rev. Lett.}\ }\textbf {\bibinfo {volume} {75}},\ \bibinfo {pages} {105}
		(\bibinfo {year} {1995})}\BibitemShut {NoStop}%
	\bibitem [{\citenamefont {Held}\ \emph {et~al.}(2001)\citenamefont {Held},
		\citenamefont {Keller}, \citenamefont {Eyert}, \citenamefont {Vollhardt},\
		and\ \citenamefont {Anisimov}}]{held;prl01}%
	\BibitemOpen
	\bibfield  {author} {\bibinfo {author} {\bibfnamefont {K.}~\bibnamefont
			{Held}}, \bibinfo {author} {\bibfnamefont {G.}~\bibnamefont {Keller}},
		\bibinfo {author} {\bibfnamefont {V.}~\bibnamefont {Eyert}}, \bibinfo
		{author} {\bibfnamefont {D.}~\bibnamefont {Vollhardt}}, \ and\ \bibinfo
		{author} {\bibfnamefont {V.~I.}\ \bibnamefont {Anisimov}},\ }\href {\doibase
		10.1103/PhysRevLett.86.5345} {\bibfield  {journal} {\bibinfo  {journal}
			{Phys. Rev. Lett.}\ }\textbf {\bibinfo {volume} {86}},\ \bibinfo {pages}
		{5345} (\bibinfo {year} {2001})}\BibitemShut {NoStop}%
	\bibitem [{\citenamefont {Mo}\ \emph {et~al.}(2003)\citenamefont {Mo},
		\citenamefont {Denlinger}, \citenamefont {Kim}, \citenamefont {Park},
		\citenamefont {Allen}, \citenamefont {Sekiyama}, \citenamefont {Yamasaki},
		\citenamefont {Kadono}, \citenamefont {Suga}, \citenamefont {Saitoh},
		\citenamefont {Muro}, \citenamefont {Metcalf}, \citenamefont {Keller},
		\citenamefont {Held}, \citenamefont {Eyert}, \citenamefont {Anisimov},\ and\
		\citenamefont {Vollhardt}}]{mo;prl03}%
	\BibitemOpen
	\bibfield  {author} {\bibinfo {author} {\bibfnamefont {S.-K.}\ \bibnamefont
			{Mo}}, \bibinfo {author} {\bibfnamefont {J.~D.}\ \bibnamefont {Denlinger}},
		\bibinfo {author} {\bibfnamefont {H.-D.}\ \bibnamefont {Kim}}, \bibinfo
		{author} {\bibfnamefont {J.-H.}\ \bibnamefont {Park}}, \bibinfo {author}
		{\bibfnamefont {J.~W.}\ \bibnamefont {Allen}}, \bibinfo {author}
		{\bibfnamefont {A.}~\bibnamefont {Sekiyama}}, \bibinfo {author}
		{\bibfnamefont {A.}~\bibnamefont {Yamasaki}}, \bibinfo {author}
		{\bibfnamefont {K.}~\bibnamefont {Kadono}}, \bibinfo {author} {\bibfnamefont
			{S.}~\bibnamefont {Suga}}, \bibinfo {author} {\bibfnamefont {Y.}~\bibnamefont
			{Saitoh}}, \bibinfo {author} {\bibfnamefont {T.}~\bibnamefont {Muro}},
		\bibinfo {author} {\bibfnamefont {P.}~\bibnamefont {Metcalf}}, \bibinfo
		{author} {\bibfnamefont {G.}~\bibnamefont {Keller}}, \bibinfo {author}
		{\bibfnamefont {K.}~\bibnamefont {Held}}, \bibinfo {author} {\bibfnamefont
			{V.}~\bibnamefont {Eyert}}, \bibinfo {author} {\bibfnamefont {V.~I.}\
			\bibnamefont {Anisimov}}, \ and\ \bibinfo {author} {\bibfnamefont
			{D.}~\bibnamefont {Vollhardt}},\ }\href {\doibase
		10.1103/PhysRevLett.90.186403} {\bibfield  {journal} {\bibinfo  {journal}
			{Phys. Rev. Lett.}\ }\textbf {\bibinfo {volume} {90}},\ \bibinfo {pages}
		{186403} (\bibinfo {year} {2003})}\BibitemShut {NoStop}%
	\bibitem [{\citenamefont {Keller}\ \emph {et~al.}(2004)\citenamefont {Keller},
		\citenamefont {Held}, \citenamefont {Eyert}, \citenamefont {Vollhardt},\ and\
		\citenamefont {Anisimov}}]{kelle;prb04}%
	\BibitemOpen
	\bibfield  {author} {\bibinfo {author} {\bibfnamefont {G.}~\bibnamefont
			{Keller}}, \bibinfo {author} {\bibfnamefont {K.}~\bibnamefont {Held}},
		\bibinfo {author} {\bibfnamefont {V.}~\bibnamefont {Eyert}}, \bibinfo
		{author} {\bibfnamefont {D.}~\bibnamefont {Vollhardt}}, \ and\ \bibinfo
		{author} {\bibfnamefont {V.~I.}\ \bibnamefont {Anisimov}},\ }\href {\doibase
		10.1103/PhysRevB.70.205116} {\bibfield  {journal} {\bibinfo  {journal} {Phys.
				Rev. B}\ }\textbf {\bibinfo {volume} {70}},\ \bibinfo {pages} {205116}
		(\bibinfo {year} {2004})}\BibitemShut {NoStop}%
	\bibitem [{\citenamefont {Kotliar}\ and\ \citenamefont
		{Vollhardt}(2004)}]{kotli;pht04}%
	\BibitemOpen
	\bibfield  {author} {\bibinfo {author} {\bibfnamefont {G.}~\bibnamefont
			{Kotliar}}\ and\ \bibinfo {author} {\bibfnamefont {D.}~\bibnamefont
			{Vollhardt}},\ }\href@noop {} {\bibfield  {journal} {\bibinfo  {journal}
			{Physics Today}\ }\textbf {\bibinfo {volume} {57}},\ \bibinfo {pages} {53}
		(\bibinfo {year} {2004})}\BibitemShut {NoStop}%
	\bibitem [{\citenamefont {Hansmann}\ \emph {et~al.}(2013)\citenamefont
		{Hansmann}, \citenamefont {Toschi}, \citenamefont {Sangiovanni},
		\citenamefont {Saha-Dasgupta}, \citenamefont {Lupi}, \citenamefont {Marsi},\
		and\ \citenamefont {Held}}]{hansm;pss13}%
	\BibitemOpen
	\bibfield  {author} {\bibinfo {author} {\bibfnamefont {P.}~\bibnamefont
			{Hansmann}}, \bibinfo {author} {\bibfnamefont {A.}~\bibnamefont {Toschi}},
		\bibinfo {author} {\bibfnamefont {G.}~\bibnamefont {Sangiovanni}}, \bibinfo
		{author} {\bibfnamefont {T.}~\bibnamefont {Saha-Dasgupta}}, \bibinfo {author}
		{\bibfnamefont {S.}~\bibnamefont {Lupi}}, \bibinfo {author} {\bibfnamefont
			{M.}~\bibnamefont {Marsi}}, \ and\ \bibinfo {author} {\bibfnamefont
			{K.}~\bibnamefont {Held}},\ }\href@noop {} {\bibfield  {journal} {\bibinfo
			{journal} {Phys. Status Solidi B}\ }\textbf {\bibinfo {volume} {250}},\
		\bibinfo {pages} {1251} (\bibinfo {year} {2013})}\BibitemShut {NoStop}%
	\bibitem [{\citenamefont {Lupi}\ \emph {et~al.}(2010)\citenamefont {Lupi},
		\citenamefont {Baldassarre}, \citenamefont {Mansart}, \citenamefont
		{Perucchi}, \citenamefont {Barinov}, \citenamefont {Dudin}, \citenamefont
		{Papalazarou}, \citenamefont {Rodolakis}, \citenamefont {Rueff},
		\citenamefont {Iti{\'{e}}}, \citenamefont {Ravy}, \citenamefont {Nicoletti},
		\citenamefont {Postorino}, \citenamefont {Hansmann}, \citenamefont {Parragh},
		\citenamefont {Toschi}, \citenamefont {Saha-Dasgupta}, \citenamefont
		{Andersen}, \citenamefont {Sangiovanni}, \citenamefont {Held},\ and\
		\citenamefont {Marsi}}]{lupi;nc10}%
	\BibitemOpen
	\bibfield  {author} {\bibinfo {author} {\bibfnamefont {S.}~\bibnamefont
			{Lupi}}, \bibinfo {author} {\bibfnamefont {L.}~\bibnamefont {Baldassarre}},
		\bibinfo {author} {\bibfnamefont {B.}~\bibnamefont {Mansart}}, \bibinfo
		{author} {\bibfnamefont {A.}~\bibnamefont {Perucchi}}, \bibinfo {author}
		{\bibfnamefont {A.}~\bibnamefont {Barinov}}, \bibinfo {author} {\bibfnamefont
			{P.}~\bibnamefont {Dudin}}, \bibinfo {author} {\bibfnamefont
			{E.}~\bibnamefont {Papalazarou}}, \bibinfo {author} {\bibfnamefont
			{F.}~\bibnamefont {Rodolakis}}, \bibinfo {author} {\bibfnamefont {J.-P.}\
			\bibnamefont {Rueff}}, \bibinfo {author} {\bibfnamefont {J.-P.}\ \bibnamefont
			{Iti{\'{e}}}}, \bibinfo {author} {\bibfnamefont {S.}~\bibnamefont {Ravy}},
		\bibinfo {author} {\bibfnamefont {D.}~\bibnamefont {Nicoletti}}, \bibinfo
		{author} {\bibfnamefont {P.}~\bibnamefont {Postorino}}, \bibinfo {author}
		{\bibfnamefont {P.}~\bibnamefont {Hansmann}}, \bibinfo {author}
		{\bibfnamefont {N.}~\bibnamefont {Parragh}}, \bibinfo {author} {\bibfnamefont
			{A.}~\bibnamefont {Toschi}}, \bibinfo {author} {\bibfnamefont
			{T.}~\bibnamefont {Saha-Dasgupta}}, \bibinfo {author} {\bibfnamefont {O.~K.}\
			\bibnamefont {Andersen}}, \bibinfo {author} {\bibfnamefont {G.}~\bibnamefont
			{Sangiovanni}}, \bibinfo {author} {\bibfnamefont {K.}~\bibnamefont {Held}}, \
		and\ \bibinfo {author} {\bibfnamefont {M.}~\bibnamefont {Marsi}},\
	}\href@noop {} {\bibfield  {journal} {\bibinfo  {journal} {Nat. Commun.}\
		}\textbf {\bibinfo {volume} {1}},\ \bibinfo {pages} {105} (\bibinfo {year}
		{2010})}\BibitemShut {NoStop}%
	\bibitem [{\citenamefont {McLeod}\ \emph {et~al.}(2016)\citenamefont {McLeod},
		\citenamefont {van Heumen}, \citenamefont {Ramirez}, \citenamefont {Wang},
		\citenamefont {Saerbeck}, \citenamefont {Guenon}, \citenamefont {Goldflam},
		\citenamefont {Anderegg}, \citenamefont {Kelly}, \citenamefont {Mueller},
		\citenamefont {Liu}, \citenamefont {Schuller},\ and\ \citenamefont
		{Basov}}]{mcleo;np16}%
	\BibitemOpen
	\bibfield  {author} {\bibinfo {author} {\bibfnamefont {A.}~\bibnamefont
			{McLeod}}, \bibinfo {author} {\bibfnamefont {E.}~\bibnamefont {van Heumen}},
		\bibinfo {author} {\bibfnamefont {J.}~\bibnamefont {Ramirez}}, \bibinfo
		{author} {\bibfnamefont {S.}~\bibnamefont {Wang}}, \bibinfo {author}
		{\bibfnamefont {T.}~\bibnamefont {Saerbeck}}, \bibinfo {author}
		{\bibfnamefont {S.}~\bibnamefont {Guenon}}, \bibinfo {author} {\bibfnamefont
			{M.}~\bibnamefont {Goldflam}}, \bibinfo {author} {\bibfnamefont
			{L.}~\bibnamefont {Anderegg}}, \bibinfo {author} {\bibfnamefont
			{P.}~\bibnamefont {Kelly}}, \bibinfo {author} {\bibfnamefont
			{A.}~\bibnamefont {Mueller}}, \bibinfo {author} {\bibfnamefont
			{M.}~\bibnamefont {Liu}}, \bibinfo {author} {\bibfnamefont {I.~K.}\
			\bibnamefont {Schuller}}, \ and\ \bibinfo {author} {\bibfnamefont
			{D.}~\bibnamefont {Basov}},\ }\href@noop {} {\bibfield  {journal} {\bibinfo
			{journal} {Nature Physics}\ }\textbf {\bibinfo {volume} {13}},\ \bibinfo
		{pages} {80} (\bibinfo {year} {2016})}\BibitemShut {NoStop}%
	\bibitem [{\citenamefont {Kalcheim}\ \emph {et~al.}(2019)\citenamefont
		{Kalcheim}, \citenamefont {Butakov}, \citenamefont {Vargas}, \citenamefont
		{Lee}, \citenamefont {del Valle}, \citenamefont {Trastoy}, \citenamefont
		{Salev}, \citenamefont {Schuller},\ and\ \citenamefont
		{Schuller}}]{kalch;prl19}%
	\BibitemOpen
	\bibfield  {author} {\bibinfo {author} {\bibfnamefont {Y.}~\bibnamefont
			{Kalcheim}}, \bibinfo {author} {\bibfnamefont {N.}~\bibnamefont {Butakov}},
		\bibinfo {author} {\bibfnamefont {N.~M.}\ \bibnamefont {Vargas}}, \bibinfo
		{author} {\bibfnamefont {M.-H.}\ \bibnamefont {Lee}}, \bibinfo {author}
		{\bibfnamefont {J.}~\bibnamefont {del Valle}}, \bibinfo {author}
		{\bibfnamefont {J.}~\bibnamefont {Trastoy}}, \bibinfo {author} {\bibfnamefont
			{P.}~\bibnamefont {Salev}}, \bibinfo {author} {\bibfnamefont
			{J.}~\bibnamefont {Schuller}}, \ and\ \bibinfo {author} {\bibfnamefont
			{I.~K.}\ \bibnamefont {Schuller}},\ }\href {\doibase
		10.1103/PhysRevLett.122.057601} {\bibfield  {journal} {\bibinfo  {journal}
			{Phys. Rev. Lett.}\ }\textbf {\bibinfo {volume} {122}},\ \bibinfo {pages}
		{057601} (\bibinfo {year} {2019})}\BibitemShut {NoStop}%
	\bibitem [{\citenamefont {Frandsen}\ \emph {et~al.}(2019)\citenamefont
		{Frandsen}, \citenamefont {Kalcheim}, \citenamefont {Valmianski},
		\citenamefont {McLeod}, \citenamefont {Guguchia}, \citenamefont {Cheung},
		\citenamefont {Hallas}, \citenamefont {Wilson}, \citenamefont {Cai},
		\citenamefont {Luke}, \citenamefont {Salman}, \citenamefont {Suter},
		\citenamefont {Prokscha}, \citenamefont {Murakami}, \citenamefont {Kageyama},
		\citenamefont {Basov}, \citenamefont {Schuller},\ and\ \citenamefont
		{Uemura}}]{frand;prb19b}%
	\BibitemOpen
	\bibfield  {author} {\bibinfo {author} {\bibfnamefont {B.~A.}\ \bibnamefont
			{Frandsen}}, \bibinfo {author} {\bibfnamefont {Y.}~\bibnamefont {Kalcheim}},
		\bibinfo {author} {\bibfnamefont {I.}~\bibnamefont {Valmianski}}, \bibinfo
		{author} {\bibfnamefont {A.~S.}\ \bibnamefont {McLeod}}, \bibinfo {author}
		{\bibfnamefont {Z.}~\bibnamefont {Guguchia}}, \bibinfo {author}
		{\bibfnamefont {S.~C.}\ \bibnamefont {Cheung}}, \bibinfo {author}
		{\bibfnamefont {A.~M.}\ \bibnamefont {Hallas}}, \bibinfo {author}
		{\bibfnamefont {M.~N.}\ \bibnamefont {Wilson}}, \bibinfo {author}
		{\bibfnamefont {Y.}~\bibnamefont {Cai}}, \bibinfo {author} {\bibfnamefont
			{G.~M.}\ \bibnamefont {Luke}}, \bibinfo {author} {\bibfnamefont
			{Z.}~\bibnamefont {Salman}}, \bibinfo {author} {\bibfnamefont
			{A.}~\bibnamefont {Suter}}, \bibinfo {author} {\bibfnamefont
			{T.}~\bibnamefont {Prokscha}}, \bibinfo {author} {\bibfnamefont
			{T.}~\bibnamefont {Murakami}}, \bibinfo {author} {\bibfnamefont
			{H.}~\bibnamefont {Kageyama}}, \bibinfo {author} {\bibfnamefont {D.~N.}\
			\bibnamefont {Basov}}, \bibinfo {author} {\bibfnamefont {I.~K.}\ \bibnamefont
			{Schuller}}, \ and\ \bibinfo {author} {\bibfnamefont {Y.~J.}\ \bibnamefont
			{Uemura}},\ }\href {\doibase 10.1103/PhysRevB.100.235136} {\bibfield
		{journal} {\bibinfo  {journal} {Phys. Rev. B}\ }\textbf {\bibinfo {volume}
			{100}},\ \bibinfo {pages} {235136} (\bibinfo {year} {2019})}\BibitemShut
	{NoStop}%
	\bibitem [{\citenamefont {Kundu}\ \emph {et~al.}(2020)\citenamefont {Kundu},
		\citenamefont {Bar}, \citenamefont {Nayak},\ and\ \citenamefont
		{Bansal}}]{kundu;prl20}%
	\BibitemOpen
	\bibfield  {author} {\bibinfo {author} {\bibfnamefont {S.}~\bibnamefont
			{Kundu}}, \bibinfo {author} {\bibfnamefont {T.}~\bibnamefont {Bar}}, \bibinfo
		{author} {\bibfnamefont {R.~K.}\ \bibnamefont {Nayak}}, \ and\ \bibinfo
		{author} {\bibfnamefont {B.}~\bibnamefont {Bansal}},\ }\href {\doibase
		10.1103/PhysRevLett.124.095703} {\bibfield  {journal} {\bibinfo  {journal}
			{Phys. Rev. Lett.}\ }\textbf {\bibinfo {volume} {124}},\ \bibinfo {pages}
		{095703} (\bibinfo {year} {2020})}\BibitemShut {NoStop}%
	\bibitem [{\citenamefont {Chen}\ \emph {et~al.}(2020)\citenamefont {Chen},
		\citenamefont {Zhou}, \citenamefont {Kalcheim}, \citenamefont {Schuller},\
		and\ \citenamefont {Natelson}}]{chen;aplm20}%
	\BibitemOpen
	\bibfield  {author} {\bibinfo {author} {\bibfnamefont {L.}~\bibnamefont
			{Chen}}, \bibinfo {author} {\bibfnamefont {P.}~\bibnamefont {Zhou}}, \bibinfo
		{author} {\bibfnamefont {Y.}~\bibnamefont {Kalcheim}}, \bibinfo {author}
		{\bibfnamefont {I.~K.}\ \bibnamefont {Schuller}}, \ and\ \bibinfo {author}
		{\bibfnamefont {D.}~\bibnamefont {Natelson}},\ }\href {\doibase
		10.1063/5.0023475} {\bibfield  {journal} {\bibinfo  {journal} {APL Mater.}\
		}\textbf {\bibinfo {volume} {8}},\ \bibinfo {pages} {101103} (\bibinfo {year}
		{2020})},\ \Eprint {http://arxiv.org/abs/https://doi.org/10.1063/5.0023475}
	{https://doi.org/10.1063/5.0023475} \BibitemShut {NoStop}%
	\bibitem [{\citenamefont {Hu}\ \emph {et~al.}(2021)\citenamefont {Hu},
		\citenamefont {Xie}, \citenamefont {Zhu}, \citenamefont {Zhu}, \citenamefont
		{Wei}, \citenamefont {Tang}, \citenamefont {Lu}, \citenamefont {Song},
		\citenamefont {Dai}, \citenamefont {Zhang}, \citenamefont {Zhang},
		\citenamefont {Zhu},\ and\ \citenamefont {Sun}}]{hu;prb21}%
	\BibitemOpen
	\bibfield  {author} {\bibinfo {author} {\bibfnamefont {L.}~\bibnamefont
			{Hu}}, \bibinfo {author} {\bibfnamefont {C.}~\bibnamefont {Xie}}, \bibinfo
		{author} {\bibfnamefont {S.~J.}\ \bibnamefont {Zhu}}, \bibinfo {author}
		{\bibfnamefont {M.}~\bibnamefont {Zhu}}, \bibinfo {author} {\bibfnamefont
			{R.~H.}\ \bibnamefont {Wei}}, \bibinfo {author} {\bibfnamefont {X.~W.}\
			\bibnamefont {Tang}}, \bibinfo {author} {\bibfnamefont {W.~J.}\ \bibnamefont
			{Lu}}, \bibinfo {author} {\bibfnamefont {W.~H.}\ \bibnamefont {Song}},
		\bibinfo {author} {\bibfnamefont {J.~M.}\ \bibnamefont {Dai}}, \bibinfo
		{author} {\bibfnamefont {R.~R.}\ \bibnamefont {Zhang}}, \bibinfo {author}
		{\bibfnamefont {C.~J.}\ \bibnamefont {Zhang}}, \bibinfo {author}
		{\bibfnamefont {X.~B.}\ \bibnamefont {Zhu}}, \ and\ \bibinfo {author}
		{\bibfnamefont {Y.~P.}\ \bibnamefont {Sun}},\ }\href {\doibase
		10.1103/PhysRevB.103.085119} {\bibfield  {journal} {\bibinfo  {journal}
			{Phys. Rev. B}\ }\textbf {\bibinfo {volume} {103}},\ \bibinfo {pages}
		{085119} (\bibinfo {year} {2021})}\BibitemShut {NoStop}%
	\bibitem [{\citenamefont {Bao}\ \emph {et~al.}(1997)\citenamefont {Bao},
		\citenamefont {Broholm}, \citenamefont {Aeppli}, \citenamefont {Dai},
		\citenamefont {Honig},\ and\ \citenamefont {Metcalf}}]{bao;prl97}%
	\BibitemOpen
	\bibfield  {author} {\bibinfo {author} {\bibfnamefont {W.}~\bibnamefont
			{Bao}}, \bibinfo {author} {\bibfnamefont {C.}~\bibnamefont {Broholm}},
		\bibinfo {author} {\bibfnamefont {G.}~\bibnamefont {Aeppli}}, \bibinfo
		{author} {\bibfnamefont {P.}~\bibnamefont {Dai}}, \bibinfo {author}
		{\bibfnamefont {J.~M.}\ \bibnamefont {Honig}}, \ and\ \bibinfo {author}
		{\bibfnamefont {P.}~\bibnamefont {Metcalf}},\ }\href {\doibase
		10.1103/PhysRevLett.78.507} {\bibfield  {journal} {\bibinfo  {journal} {Phys.
				Rev. Lett.}\ }\textbf {\bibinfo {volume} {78}},\ \bibinfo {pages} {507}
		(\bibinfo {year} {1997})}\BibitemShut {NoStop}%
	\bibitem [{\citenamefont {Bao}\ \emph {et~al.}(1998)\citenamefont {Bao},
		\citenamefont {Broholm}, \citenamefont {Aeppli}, \citenamefont {Carter},
		\citenamefont {Dai}, \citenamefont {Rosenbaum}, \citenamefont {Honig},
		\citenamefont {Metcalf},\ and\ \citenamefont {Trevino}}]{bao;prb98}%
	\BibitemOpen
	\bibfield  {author} {\bibinfo {author} {\bibfnamefont {W.}~\bibnamefont
			{Bao}}, \bibinfo {author} {\bibfnamefont {C.}~\bibnamefont {Broholm}},
		\bibinfo {author} {\bibfnamefont {G.}~\bibnamefont {Aeppli}}, \bibinfo
		{author} {\bibfnamefont {S.~A.}\ \bibnamefont {Carter}}, \bibinfo {author}
		{\bibfnamefont {P.}~\bibnamefont {Dai}}, \bibinfo {author} {\bibfnamefont
			{T.~F.}\ \bibnamefont {Rosenbaum}}, \bibinfo {author} {\bibfnamefont {J.~M.}\
			\bibnamefont {Honig}}, \bibinfo {author} {\bibfnamefont {P.}~\bibnamefont
			{Metcalf}}, \ and\ \bibinfo {author} {\bibfnamefont {S.~F.}\ \bibnamefont
			{Trevino}},\ }\href {\doibase 10.1103/PhysRevB.58.12727} {\bibfield
		{journal} {\bibinfo  {journal} {Phys. Rev. B}\ }\textbf {\bibinfo {volume}
			{58}},\ \bibinfo {pages} {12727} (\bibinfo {year} {1998})}\BibitemShut
	{NoStop}%
	\bibitem [{\citenamefont {Trastoy}\ \emph {et~al.}(2020)\citenamefont
		{Trastoy}, \citenamefont {Camjayi}, \citenamefont {del Valle}, \citenamefont
		{Kalcheim}, \citenamefont {Crocombette}, \citenamefont {Gilbert},
		\citenamefont {Borchers}, \citenamefont {Villegas}, \citenamefont
		{Ravelosona}, \citenamefont {Rozenberg},\ and\ \citenamefont
		{Schuller}}]{trast;prb20}%
	\BibitemOpen
	\bibfield  {author} {\bibinfo {author} {\bibfnamefont {J.}~\bibnamefont
			{Trastoy}}, \bibinfo {author} {\bibfnamefont {A.}~\bibnamefont {Camjayi}},
		\bibinfo {author} {\bibfnamefont {J.}~\bibnamefont {del Valle}}, \bibinfo
		{author} {\bibfnamefont {Y.}~\bibnamefont {Kalcheim}}, \bibinfo {author}
		{\bibfnamefont {J.-P.}\ \bibnamefont {Crocombette}}, \bibinfo {author}
		{\bibfnamefont {D.~A.}\ \bibnamefont {Gilbert}}, \bibinfo {author}
		{\bibfnamefont {J.~A.}\ \bibnamefont {Borchers}}, \bibinfo {author}
		{\bibfnamefont {J.~E.}\ \bibnamefont {Villegas}}, \bibinfo {author}
		{\bibfnamefont {D.}~\bibnamefont {Ravelosona}}, \bibinfo {author}
		{\bibfnamefont {M.~J.}\ \bibnamefont {Rozenberg}}, \ and\ \bibinfo {author}
		{\bibfnamefont {I.~K.}\ \bibnamefont {Schuller}},\ }\href {\doibase
		10.1103/PhysRevB.101.245109} {\bibfield  {journal} {\bibinfo  {journal}
			{Phys. Rev. B}\ }\textbf {\bibinfo {volume} {101}},\ \bibinfo {pages}
		{245109} (\bibinfo {year} {2020})}\BibitemShut {NoStop}%
	\bibitem [{\citenamefont {Frenkel}\ \emph {et~al.}(1997)\citenamefont
		{Frenkel}, \citenamefont {Stern},\ and\ \citenamefont
		{Chudnovsky}}]{frenk;ssc97}%
	\BibitemOpen
	\bibfield  {author} {\bibinfo {author} {\bibfnamefont {A.~I.}\ \bibnamefont
			{Frenkel}}, \bibinfo {author} {\bibfnamefont {E.~A.}\ \bibnamefont {Stern}},
		\ and\ \bibinfo {author} {\bibfnamefont {F.~A.}\ \bibnamefont {Chudnovsky}},\
	}\href@noop {} {\bibfield  {journal} {\bibinfo  {journal} {Solid State
				Commun.}\ }\textbf {\bibinfo {volume} {102}},\ \bibinfo {pages} {637}
		(\bibinfo {year} {1997})}\BibitemShut {NoStop}%
	\bibitem [{\citenamefont {Pfalzer}\ \emph {et~al.}(2006)\citenamefont
		{Pfalzer}, \citenamefont {Obermeier}, \citenamefont {Klemm}, \citenamefont
		{Horn},\ and\ \citenamefont {denBoer}}]{pfalz;prb06}%
	\BibitemOpen
	\bibfield  {author} {\bibinfo {author} {\bibfnamefont {P.}~\bibnamefont
			{Pfalzer}}, \bibinfo {author} {\bibfnamefont {G.}~\bibnamefont {Obermeier}},
		\bibinfo {author} {\bibfnamefont {M.}~\bibnamefont {Klemm}}, \bibinfo
		{author} {\bibfnamefont {S.}~\bibnamefont {Horn}}, \ and\ \bibinfo {author}
		{\bibfnamefont {M.~L.}\ \bibnamefont {denBoer}},\ }\href {\doibase
		10.1103/PhysRevB.73.144106} {\bibfield  {journal} {\bibinfo  {journal} {Phys.
				Rev. B}\ }\textbf {\bibinfo {volume} {73}},\ \bibinfo {pages} {144106}
		(\bibinfo {year} {2006})}\BibitemShut {NoStop}%
	\bibitem [{\citenamefont {M\"uller}\ \emph {et~al.}(2005)\citenamefont
		{M\"uller}, \citenamefont {Nateprov}, \citenamefont {Obermeier},
		\citenamefont {Klemm}, \citenamefont {Tidecks}, \citenamefont {Wixforth},\
		and\ \citenamefont {Horn}}]{mulle;jap05}%
	\BibitemOpen
	\bibfield  {author} {\bibinfo {author} {\bibfnamefont {C.}~\bibnamefont
			{M\"uller}}, \bibinfo {author} {\bibfnamefont {A.~A.}\ \bibnamefont
			{Nateprov}}, \bibinfo {author} {\bibfnamefont {G.}~\bibnamefont {Obermeier}},
		\bibinfo {author} {\bibfnamefont {M.}~\bibnamefont {Klemm}}, \bibinfo
		{author} {\bibfnamefont {R.}~\bibnamefont {Tidecks}}, \bibinfo {author}
		{\bibfnamefont {A.}~\bibnamefont {Wixforth}}, \ and\ \bibinfo {author}
		{\bibfnamefont {S.}~\bibnamefont {Horn}},\ }\href {\doibase
		10.1063/1.2103410} {\bibfield  {journal} {\bibinfo  {journal} {J. Appl.
				Phys.}\ }\textbf {\bibinfo {volume} {98}},\ \bibinfo {pages} {084111}
		(\bibinfo {year} {2005})}\BibitemShut {NoStop}%
	\bibitem [{\citenamefont {M\"uller}\ \emph {et~al.}(2008)\citenamefont
		{M\"uller}, \citenamefont {Nateprov}, \citenamefont {Klemm}, \citenamefont
		{Wixforth}, \citenamefont {Tidecks},\ and\ \citenamefont
		{Horn}}]{mulle;jap08}%
	\BibitemOpen
	\bibfield  {author} {\bibinfo {author} {\bibfnamefont {C.}~\bibnamefont
			{M\"uller}}, \bibinfo {author} {\bibfnamefont {A.~A.}\ \bibnamefont
			{Nateprov}}, \bibinfo {author} {\bibfnamefont {M.}~\bibnamefont {Klemm}},
		\bibinfo {author} {\bibfnamefont {A.}~\bibnamefont {Wixforth}}, \bibinfo
		{author} {\bibfnamefont {R.}~\bibnamefont {Tidecks}}, \ and\ \bibinfo
		{author} {\bibfnamefont {S.}~\bibnamefont {Horn}},\ }\href {\doibase
		10.1063/1.2871302} {\bibfield  {journal} {\bibinfo  {journal} {J. Appl.
				Phys.}\ }\textbf {\bibinfo {volume} {103}},\ \bibinfo {pages} {063705}
		(\bibinfo {year} {2008})}\BibitemShut {NoStop}%
	\bibitem [{\citenamefont {K{\"u}ndel}\ \emph {et~al.}(2013)\citenamefont
		{K{\"u}ndel}, \citenamefont {Pontiller}, \citenamefont {Müller},
		\citenamefont {Obermeier}, \citenamefont {Liu}, \citenamefont {Nateprov},
		\citenamefont {Hörner}, \citenamefont {Wixforth}, \citenamefont {Horn},\
		and\ \citenamefont {Tidecks}}]{kunde;apl13}%
	\BibitemOpen
	\bibfield  {author} {\bibinfo {author} {\bibfnamefont {J.}~\bibnamefont
			{K{\"u}ndel}}, \bibinfo {author} {\bibfnamefont {P.}~\bibnamefont
			{Pontiller}}, \bibinfo {author} {\bibfnamefont {C.}~\bibnamefont {Müller}},
		\bibinfo {author} {\bibfnamefont {G.}~\bibnamefont {Obermeier}}, \bibinfo
		{author} {\bibfnamefont {Z.}~\bibnamefont {Liu}}, \bibinfo {author}
		{\bibfnamefont {A.~A.}\ \bibnamefont {Nateprov}}, \bibinfo {author}
		{\bibfnamefont {A.}~\bibnamefont {Hörner}}, \bibinfo {author} {\bibfnamefont
			{A.}~\bibnamefont {Wixforth}}, \bibinfo {author} {\bibfnamefont
			{S.}~\bibnamefont {Horn}}, \ and\ \bibinfo {author} {\bibfnamefont
			{R.}~\bibnamefont {Tidecks}},\ }\href {\doibase 10.1063/1.4794948} {\bibfield
		{journal} {\bibinfo  {journal} {Appl. Phys. Lett.}\ }\textbf {\bibinfo
			{volume} {102}},\ \bibinfo {pages} {101904} (\bibinfo {year}
		{2013})}\BibitemShut {NoStop}%
	\bibitem [{\citenamefont {Seikh}\ \emph {et~al.}(2006)\citenamefont {Seikh},
		\citenamefont {Narayana}, \citenamefont {Sood}, \citenamefont {Murugavel},
		\citenamefont {Kim}, \citenamefont {Metcalf}, \citenamefont {Honig},\ and\
		\citenamefont {Rao}}]{seikh;ssc06}%
	\BibitemOpen
	\bibfield  {author} {\bibinfo {author} {\bibfnamefont {M.~M.}\ \bibnamefont
			{Seikh}}, \bibinfo {author} {\bibfnamefont {C.}~\bibnamefont {Narayana}},
		\bibinfo {author} {\bibfnamefont {A.~K.}\ \bibnamefont {Sood}}, \bibinfo
		{author} {\bibfnamefont {P.}~\bibnamefont {Murugavel}}, \bibinfo {author}
		{\bibfnamefont {M.~W.}\ \bibnamefont {Kim}}, \bibinfo {author} {\bibfnamefont
			{P.~A.}\ \bibnamefont {Metcalf}}, \bibinfo {author} {\bibfnamefont {J.~M.}\
			\bibnamefont {Honig}}, \ and\ \bibinfo {author} {\bibfnamefont {C.~N.~R.}\
			\bibnamefont {Rao}},\ }\href {\doibase 10.1016/j.ssc.2006.03.026} {\bibfield
		{journal} {\bibinfo  {journal} {Solid State Commun.}\ }\textbf {\bibinfo
			{volume} {138}},\ \bibinfo {pages} {466} (\bibinfo {year}
		{2006})}\BibitemShut {NoStop}%
	\bibitem [{\citenamefont {Billinge}\ \emph {et~al.}(1996)\citenamefont
		{Billinge}, \citenamefont {DiFrancesco}, \citenamefont {Kwei}, \citenamefont
		{Neumeier},\ and\ \citenamefont {Thompson}}]{billi;prl96}%
	\BibitemOpen
	\bibfield  {author} {\bibinfo {author} {\bibfnamefont {S.~J.~L.}\
			\bibnamefont {Billinge}}, \bibinfo {author} {\bibfnamefont {R.~G.}\
			\bibnamefont {DiFrancesco}}, \bibinfo {author} {\bibfnamefont {G.~H.}\
			\bibnamefont {Kwei}}, \bibinfo {author} {\bibfnamefont {J.~J.}\ \bibnamefont
			{Neumeier}}, \ and\ \bibinfo {author} {\bibfnamefont {J.~D.}\ \bibnamefont
			{Thompson}},\ }\href {\doibase 10.1103/PhysRevLett.77.715} {\bibfield
		{journal} {\bibinfo  {journal} {Phys. Rev. Lett.}\ }\textbf {\bibinfo
			{volume} {77}},\ \bibinfo {pages} {715} (\bibinfo {year} {1996})}\BibitemShut
	{NoStop}%
	\bibitem [{\citenamefont {Qiu}\ \emph {et~al.}(2005)\citenamefont {Qiu},
		\citenamefont {Proffen}, \citenamefont {Mitchell},\ and\ \citenamefont
		{Billinge}}]{qiu;prl05}%
	\BibitemOpen
	\bibfield  {author} {\bibinfo {author} {\bibfnamefont {X.}~\bibnamefont
			{Qiu}}, \bibinfo {author} {\bibfnamefont {T.}~\bibnamefont {Proffen}},
		\bibinfo {author} {\bibfnamefont {J.~F.}\ \bibnamefont {Mitchell}}, \ and\
		\bibinfo {author} {\bibfnamefont {S.~J.~L.}\ \bibnamefont {Billinge}},\
	}\href {http://journals.aps.org/prl/abstract/10.1103/PhysRevLett.94.177203}
	{\bibfield  {journal} {\bibinfo  {journal} {Phys. Rev. Lett.}\ }\textbf
		{\bibinfo {volume} {94}},\ \bibinfo {pages} {177203} (\bibinfo {year}
		{2005})}\BibitemShut {NoStop}%
	\bibitem [{\citenamefont {Dagotto}(2005)}]{dagot;s05}%
	\BibitemOpen
	\bibfield  {author} {\bibinfo {author} {\bibfnamefont {E.}~\bibnamefont
			{Dagotto}},\ }\href {\doibase 10.1126/science.1107559} {\bibfield  {journal}
		{\bibinfo  {journal} {Science}\ }\textbf {\bibinfo {volume} {309}},\ \bibinfo
		{pages} {257} (\bibinfo {year} {2005})}\BibitemShut {NoStop}%
	\bibitem [{\citenamefont {Bo\v{z}in}\ \emph {et~al.}(2014)\citenamefont
		{Bo\v{z}in}, \citenamefont {Knox}, \citenamefont {Juh\'{a}s}, \citenamefont
		{Hor}, \citenamefont {Mitchell},\ and\ \citenamefont
		{Billinge}}]{bozin;sr14}%
	\BibitemOpen
	\bibfield  {author} {\bibinfo {author} {\bibfnamefont {E.~S.}\ \bibnamefont
			{Bo\v{z}in}}, \bibinfo {author} {\bibfnamefont {K.~R.}\ \bibnamefont {Knox}},
		\bibinfo {author} {\bibfnamefont {P.}~\bibnamefont {Juh\'{a}s}}, \bibinfo
		{author} {\bibfnamefont {Y.~S.}\ \bibnamefont {Hor}}, \bibinfo {author}
		{\bibfnamefont {J.~F.}\ \bibnamefont {Mitchell}}, \ and\ \bibinfo {author}
		{\bibfnamefont {S.~J.~L.}\ \bibnamefont {Billinge}},\ }\href {\doibase
		10.1038/srep04081} {\bibfield  {journal} {\bibinfo  {journal} {Sci. Rep.}\
		}\textbf {\bibinfo {volume} {4}},\ \bibinfo {pages} {4081} (\bibinfo {year}
		{2014})}\BibitemShut {NoStop}%
	\bibitem [{\citenamefont {Bozin}\ \emph {et~al.}(2019)\citenamefont {Bozin},
		\citenamefont {Yin}, \citenamefont {Koch}, \citenamefont {Abeykoon},
		\citenamefont {Hor}, \citenamefont {Zheng}, \citenamefont {Lei},
		\citenamefont {Petrovic}, \citenamefont {Mitchell},\ and\ \citenamefont
		{Billinge}}]{bozin;nc19}%
	\BibitemOpen
	\bibfield  {author} {\bibinfo {author} {\bibfnamefont {E.~S.}\ \bibnamefont
			{Bozin}}, \bibinfo {author} {\bibfnamefont {W.~G.}\ \bibnamefont {Yin}},
		\bibinfo {author} {\bibfnamefont {R.~J.}\ \bibnamefont {Koch}}, \bibinfo
		{author} {\bibfnamefont {M.}~\bibnamefont {Abeykoon}}, \bibinfo {author}
		{\bibfnamefont {Y.~S.}\ \bibnamefont {Hor}}, \bibinfo {author} {\bibfnamefont
			{H.}~\bibnamefont {Zheng}}, \bibinfo {author} {\bibfnamefont {H.~C.}\
			\bibnamefont {Lei}}, \bibinfo {author} {\bibfnamefont {C.}~\bibnamefont
			{Petrovic}}, \bibinfo {author} {\bibfnamefont {J.~F.}\ \bibnamefont
			{Mitchell}}, \ and\ \bibinfo {author} {\bibfnamefont {S.~J.~L.}\ \bibnamefont
			{Billinge}},\ }\href@noop {} {\bibfield  {journal} {\bibinfo  {journal} {Nat.
				Commun.}\ }\textbf {\bibinfo {volume} {10}},\ \bibinfo {pages} {3638}
		(\bibinfo {year} {2019})}\BibitemShut {NoStop}%
	\bibitem [{\citenamefont {Perversi}\ \emph {et~al.}(2019)\citenamefont
		{Perversi}, \citenamefont {Pachoud}, \citenamefont {Cumby}, \citenamefont
		{Hudspeth}, \citenamefont {Wright}, \citenamefont {Kimber},\ and\
		\citenamefont {Attfield}}]{perve;nc19}%
	\BibitemOpen
	\bibfield  {author} {\bibinfo {author} {\bibfnamefont {G.}~\bibnamefont
			{Perversi}}, \bibinfo {author} {\bibfnamefont {E.}~\bibnamefont {Pachoud}},
		\bibinfo {author} {\bibfnamefont {J.}~\bibnamefont {Cumby}}, \bibinfo
		{author} {\bibfnamefont {J.~M.}\ \bibnamefont {Hudspeth}}, \bibinfo {author}
		{\bibfnamefont {J.~P.}\ \bibnamefont {Wright}}, \bibinfo {author}
		{\bibfnamefont {S.~A.~J.}\ \bibnamefont {Kimber}}, \ and\ \bibinfo {author}
		{\bibfnamefont {J.~P.}\ \bibnamefont {Attfield}},\ }\href {\doibase
		10.1038/s41467-019-10949-9} {\bibfield  {journal} {\bibinfo  {journal} {Nat.
				Commun.}\ }\textbf {\bibinfo {volume} {10}},\ \bibinfo {pages} {2857}
		(\bibinfo {year} {2019})}\BibitemShut {NoStop}%
	\bibitem [{\citenamefont {Egami}\ and\ \citenamefont
		{Billinge}(2012)}]{egami;b;utbp12}%
	\BibitemOpen
	\bibfield  {author} {\bibinfo {author} {\bibfnamefont {T.}~\bibnamefont
			{Egami}}\ and\ \bibinfo {author} {\bibfnamefont {S.~J.~L.}\ \bibnamefont
			{Billinge}},\ }\href
	{http://store.elsevier.com/product.jsp?lid=0\&iid=73\&sid=0\&isbn=9780080971414}
	{\emph {\bibinfo {title} {Underneath the Bragg peaks: structural analysis of
				complex materials}}},\ \bibinfo {edition} {2nd}\ ed.\ (\bibinfo  {publisher}
	{Elsevier},\ \bibinfo {address} {Amsterdam},\ \bibinfo {year}
	{2012})\BibitemShut {NoStop}%
	\bibitem [{\citenamefont {Neuefeind}\ \emph {et~al.}(2012)\citenamefont
		{Neuefeind}, \citenamefont {Feygenson}, \citenamefont {Carruth},
		\citenamefont {Hoffmann},\ and\ \citenamefont {Chipley}}]{neuef;nimb12}%
	\BibitemOpen
	\bibfield  {author} {\bibinfo {author} {\bibfnamefont {J.}~\bibnamefont
			{Neuefeind}}, \bibinfo {author} {\bibfnamefont {M.}~\bibnamefont
			{Feygenson}}, \bibinfo {author} {\bibfnamefont {J.}~\bibnamefont {Carruth}},
		\bibinfo {author} {\bibfnamefont {R.}~\bibnamefont {Hoffmann}}, \ and\
		\bibinfo {author} {\bibfnamefont {K.~K.}\ \bibnamefont {Chipley}},\ }\href
	{\doibase 10.1016/j.nimb.2012.05.037} {\bibfield  {journal} {\bibinfo
			{journal} {Nucl. Instrum. Meth. B}\ }\textbf {\bibinfo {volume} {287}},\
		\bibinfo {pages} {68 } (\bibinfo {year} {2012})}\BibitemShut {NoStop}%
	\bibitem [{\citenamefont {Hammersley}(2004)}]{hamme;esrf04}%
	\BibitemOpen
	\bibfield  {author} {\bibinfo {author} {\bibfnamefont {A.~P.}\ \bibnamefont
			{Hammersley}},\ }\href@noop {} {\enquote {\bibinfo {title} {Fit2d v12. 012
				reference manual v6.0},}\ } (\bibinfo {year} {2004}),\ \bibinfo {note}
	{{ESRF} Internal Report {ESRF98HA01T}}\BibitemShut {NoStop}%
	\bibitem [{\citenamefont {Farrow}\ \emph {et~al.}(2007)\citenamefont {Farrow},
		\citenamefont {Juh\'as}, \citenamefont {Liu}, \citenamefont {Bryndin},
		\citenamefont {{Bo\v zin}}, \citenamefont {Bloch}, \citenamefont {Proffen},\
		and\ \citenamefont {Billinge}}]{farro;jpcm07}%
	\BibitemOpen
	\bibfield  {author} {\bibinfo {author} {\bibfnamefont {C.~L.}\ \bibnamefont
			{Farrow}}, \bibinfo {author} {\bibfnamefont {P.}~\bibnamefont {Juh\'as}},
		\bibinfo {author} {\bibfnamefont {J.}~\bibnamefont {Liu}}, \bibinfo {author}
		{\bibfnamefont {D.}~\bibnamefont {Bryndin}}, \bibinfo {author} {\bibfnamefont
			{E.~S.}\ \bibnamefont {{Bo\v zin}}}, \bibinfo {author} {\bibfnamefont
			{J.}~\bibnamefont {Bloch}}, \bibinfo {author} {\bibfnamefont
			{T.}~\bibnamefont {Proffen}}, \ and\ \bibinfo {author} {\bibfnamefont
			{S.~J.~L.}\ \bibnamefont {Billinge}},\ }\href {\doibase
		10.1088/0953-8984/19/33/335219} {\bibfield  {journal} {\bibinfo  {journal}
			{J. Phys.: Condens. Mat.}\ }\textbf {\bibinfo {volume} {19}},\ \bibinfo
		{pages} {335219} (\bibinfo {year} {2007})}\BibitemShut {NoStop}%
	\bibitem [{\citenamefont {Juh\'{a}s}\ \emph {et~al.}(2015)\citenamefont
		{Juh\'{a}s}, \citenamefont {Farrow}, \citenamefont {Yang}, \citenamefont
		{Knox},\ and\ \citenamefont {Billinge}}]{juhas;aca15}%
	\BibitemOpen
	\bibfield  {author} {\bibinfo {author} {\bibfnamefont {P.}~\bibnamefont
			{Juh\'{a}s}}, \bibinfo {author} {\bibfnamefont {C.~L.}\ \bibnamefont
			{Farrow}}, \bibinfo {author} {\bibfnamefont {X.}~\bibnamefont {Yang}},
		\bibinfo {author} {\bibfnamefont {K.~R.}\ \bibnamefont {Knox}}, \ and\
		\bibinfo {author} {\bibfnamefont {S.~J.~L.}\ \bibnamefont {Billinge}},\
	}\href {\doibase 10.1107/S2053273315014473} {\bibfield  {journal} {\bibinfo
			{journal} {Acta Crystallogr. A}\ }\textbf {\bibinfo {volume} {71}},\ \bibinfo
		{pages} {562} (\bibinfo {year} {2015})}\BibitemShut {NoStop}%
	\bibitem [{\citenamefont {Frandsen}\ \emph {et~al.}(2014)\citenamefont
		{Frandsen}, \citenamefont {Yang},\ and\ \citenamefont
		{Billinge}}]{frand;aca14}%
	\BibitemOpen
	\bibfield  {author} {\bibinfo {author} {\bibfnamefont {B.~A.}\ \bibnamefont
			{Frandsen}}, \bibinfo {author} {\bibfnamefont {X.}~\bibnamefont {Yang}}, \
		and\ \bibinfo {author} {\bibfnamefont {S.~J.~L.}\ \bibnamefont {Billinge}},\
	}\href {\doibase 10.1107/S2053273313033081} {\bibfield  {journal} {\bibinfo
			{journal} {Acta Crystallogr. A}\ }\textbf {\bibinfo {volume} {70}},\ \bibinfo
		{pages} {3} (\bibinfo {year} {2014})}\BibitemShut {NoStop}%
	\bibitem [{\citenamefont {Frandsen}\ and\ \citenamefont
		{Billinge}(2015)}]{frand;aca15}%
	\BibitemOpen
	\bibfield  {author} {\bibinfo {author} {\bibfnamefont {B.~A.}\ \bibnamefont
			{Frandsen}}\ and\ \bibinfo {author} {\bibfnamefont {S.~J.~L.}\ \bibnamefont
			{Billinge}},\ }\href {\doibase 10.1107/S205327331500306X} {\bibfield
		{journal} {\bibinfo  {journal} {Acta Crystallogr. A}\ }\textbf {\bibinfo
			{volume} {71}},\ \bibinfo {pages} {325} (\bibinfo {year} {2015})}\BibitemShut
	{NoStop}%
	\bibitem [{\citenamefont {Frandsen}\ \emph
		{et~al.}(2016{\natexlab{b}})\citenamefont {Frandsen}, \citenamefont
		{Brunelli}, \citenamefont {Page}, \citenamefont {Uemura}, \citenamefont
		{Staunton},\ and\ \citenamefont {Billinge}}]{frand;prl16}%
	\BibitemOpen
	\bibfield  {author} {\bibinfo {author} {\bibfnamefont {B.~A.}\ \bibnamefont
			{Frandsen}}, \bibinfo {author} {\bibfnamefont {M.}~\bibnamefont {Brunelli}},
		\bibinfo {author} {\bibfnamefont {K.}~\bibnamefont {Page}}, \bibinfo {author}
		{\bibfnamefont {Y.~J.}\ \bibnamefont {Uemura}}, \bibinfo {author}
		{\bibfnamefont {J.~B.}\ \bibnamefont {Staunton}}, \ and\ \bibinfo {author}
		{\bibfnamefont {S.~J.~L.}\ \bibnamefont {Billinge}},\ }\href {\doibase
		10.1103/PhysRevLett.116.197204} {\bibfield  {journal} {\bibinfo  {journal}
			{Phys. Rev. Lett.}\ }\textbf {\bibinfo {volume} {116}},\ \bibinfo {pages}
		{197204} (\bibinfo {year} {2016}{\natexlab{b}})}\BibitemShut {NoStop}%
	\bibitem [{\citenamefont {Bird}\ \emph {et~al.}(2021)\citenamefont {Bird},
		\citenamefont {Herlihy},\ and\ \citenamefont {Senn}}]{bird;jac21}%
	\BibitemOpen
	\bibfield  {author} {\bibinfo {author} {\bibfnamefont {T.~A.}\ \bibnamefont
			{Bird}}, \bibinfo {author} {\bibfnamefont {A.}~\bibnamefont {Herlihy}}, \
		and\ \bibinfo {author} {\bibfnamefont {M.~S.}\ \bibnamefont {Senn}},\ }\href
	{\doibase 10.1107/S1600576721008499} {\bibfield  {journal} {\bibinfo
			{journal} {J. Appl. Crystallogr.}\ }\textbf {\bibinfo {volume} {54}},\
		\bibinfo {pages} {1514} (\bibinfo {year} {2021})}\BibitemShut {NoStop}%
	\bibitem [{\citenamefont {Frandsen}\ \emph {et~al.}(2020)\citenamefont
		{Frandsen}, \citenamefont {Bozin}, \citenamefont {Aza}, \citenamefont
		{Mart\'{\i}nez}, \citenamefont {Feygenson}, \citenamefont {Page},\ and\
		\citenamefont {Lappas}}]{frand;prb20}%
	\BibitemOpen
	\bibfield  {author} {\bibinfo {author} {\bibfnamefont {B.~A.}\ \bibnamefont
			{Frandsen}}, \bibinfo {author} {\bibfnamefont {E.~S.}\ \bibnamefont {Bozin}},
		\bibinfo {author} {\bibfnamefont {E.}~\bibnamefont {Aza}}, \bibinfo {author}
		{\bibfnamefont {A.~F.}\ \bibnamefont {Mart\'{\i}nez}}, \bibinfo {author}
		{\bibfnamefont {M.}~\bibnamefont {Feygenson}}, \bibinfo {author}
		{\bibfnamefont {K.}~\bibnamefont {Page}}, \ and\ \bibinfo {author}
		{\bibfnamefont {A.}~\bibnamefont {Lappas}},\ }\href {\doibase
		10.1103/PhysRevB.101.024423} {\bibfield  {journal} {\bibinfo  {journal}
			{Phys. Rev. B}\ }\textbf {\bibinfo {volume} {101}},\ \bibinfo {pages}
		{024423} (\bibinfo {year} {2020})}\BibitemShut {NoStop}%
	\bibitem [{\citenamefont {Frandsen}\ \emph {et~al.}(2017)\citenamefont
		{Frandsen}, \citenamefont {Ross}, \citenamefont {Krizan}, \citenamefont
		{Nilsen}, \citenamefont {Wildes}, \citenamefont {Cava}, \citenamefont
		{Birgeneau},\ and\ \citenamefont {Billinge}}]{frand;prm17}%
	\BibitemOpen
	\bibfield  {author} {\bibinfo {author} {\bibfnamefont {B.~A.}\ \bibnamefont
			{Frandsen}}, \bibinfo {author} {\bibfnamefont {K.~A.}\ \bibnamefont {Ross}},
		\bibinfo {author} {\bibfnamefont {J.~W.}\ \bibnamefont {Krizan}}, \bibinfo
		{author} {\bibfnamefont {G.~J.}\ \bibnamefont {Nilsen}}, \bibinfo {author}
		{\bibfnamefont {A.~R.}\ \bibnamefont {Wildes}}, \bibinfo {author}
		{\bibfnamefont {R.~J.}\ \bibnamefont {Cava}}, \bibinfo {author}
		{\bibfnamefont {R.~J.}\ \bibnamefont {Birgeneau}}, \ and\ \bibinfo {author}
		{\bibfnamefont {S.~J.~L.}\ \bibnamefont {Billinge}},\ }\href {\doibase
		10.1103/PhysRevMaterials.1.074412} {\bibfield  {journal} {\bibinfo  {journal}
			{Phys. Rev. Materials}\ }\textbf {\bibinfo {volume} {1}},\ \bibinfo {pages}
		{074412} (\bibinfo {year} {2017})}\BibitemShut {NoStop}%
	\bibitem [{\citenamefont {Kodama}\ \emph {et~al.}(2017)\citenamefont {Kodama},
		\citenamefont {Ikeda}, \citenamefont {Shamoto},\ and\ \citenamefont
		{Otomo}}]{kodam;jpsj17}%
	\BibitemOpen
	\bibfield  {author} {\bibinfo {author} {\bibfnamefont {K.}~\bibnamefont
			{Kodama}}, \bibinfo {author} {\bibfnamefont {K.}~\bibnamefont {Ikeda}},
		\bibinfo {author} {\bibfnamefont {S.-i.}\ \bibnamefont {Shamoto}}, \ and\
		\bibinfo {author} {\bibfnamefont {T.}~\bibnamefont {Otomo}},\ }\href
	{\doibase 10.7566/JPSJ.86.124708} {\bibfield  {journal} {\bibinfo  {journal}
			{J. Phys. Soc. Jpn}\ }\textbf {\bibinfo {volume} {86}},\ \bibinfo {pages}
		{124708} (\bibinfo {year} {2017})}\BibitemShut {NoStop}%
	\bibitem [{\citenamefont {Zhang}\ \emph {et~al.}(2019)\citenamefont {Zhang},
		\citenamefont {Scholz}, \citenamefont {Dronskowski}, \citenamefont
		{McDonnell},\ and\ \citenamefont {Tucker}}]{zhang;prb19}%
	\BibitemOpen
	\bibfield  {author} {\bibinfo {author} {\bibfnamefont {Y.}~\bibnamefont
			{Zhang}}, \bibinfo {author} {\bibfnamefont {T.}~\bibnamefont {Scholz}},
		\bibinfo {author} {\bibfnamefont {R.}~\bibnamefont {Dronskowski}}, \bibinfo
		{author} {\bibfnamefont {M.~T.}\ \bibnamefont {McDonnell}}, \ and\ \bibinfo
		{author} {\bibfnamefont {M.~G.}\ \bibnamefont {Tucker}},\ }\href {\doibase
		10.1103/PhysRevB.100.014419} {\bibfield  {journal} {\bibinfo  {journal}
			{Phys. Rev. B}\ }\textbf {\bibinfo {volume} {100}},\ \bibinfo {pages}
		{014419} (\bibinfo {year} {2019})}\BibitemShut {NoStop}%
\end{thebibliography}
\end{document}